\documentclass[9pt,twocolumn,twoside]{osajnl}

\journal{jocn} 

\setboolean{shortarticle}{false}
\title{RADICAL: Reconfiguration-Aware Direct-Connect AI Cluster using Spatial-and-Wavelength Selective Switching}

\author[1, *]{Brett George}
\author[2]{Weiyang Wang}
\author[1,3]{Yuyang Wang}
\author[1,4]{Zhenguo Wu}
\author[5]{S. C. Mancas}
\author[1]{Xiang Meng}
\author[2,6]{Manya Ghobadi}
\author[1]{Keren Bergman}

\affil[1]{Department of Electrical Engineering, Columbia University, New York, NY 10027, USA
}
\affil[2]{Computer Science and Artificial Intelligence Laboratory, MIT, Cambridge, MA 02139, USA}
\affil[3]{Now with Dept. of Electrical and Computer Engineering, University of Connecticut, Storrs, CT 06269, USA}
\affil[4]{Now with NVIDIA Corporation, Santa Clara, CA 95051, USA}
\affil[5]{Laboratory for Physical Sciences, 5520 Research Park Drive, Catonsville, MD 21228, USA}
\affil[6]{Systalyze, 1 Broadway, Cambridge, MA 02142, USA}

\affil[*]{brett.george@columbia.edu}

\begin{abstract}
We present RADICAL, the silicon-photonic-enabled \underline{R}econfiguration-\underline{A}ware \underline{Di}rect-\underline{C}onnect \underline{A}I C\underline{l}uster that integrates a custom spatial-and-wavelength selective switch (SWSS) with the Linux network stack. RADICAL deploys a packaged $4\times4\times4\lambda$ microring-based SWSS into a four-GPU-server cluster using standard SFP28 optical transceivers, and a reconfiguration-aware host-networking interface to demonstrate a direct-connect reconfigurable optical fabric. The SWSS provides dynamic spatial and wavelength routing with $3.2-6.8$ dB path loss, average wavelength and spatial crosstalk isolation exceeding $25$ dB, and error-free $25$ Gbps signal recovery across routing paths. At the system level, RADICAL validates reconfiguration-aware operation through continuous bulk data transfers totaling 6.4 TBytes over 10 minutes. We further demonstrate multitenant AI training across reconfigurable link topologies, including four ResNet-18 instances and six nanoGPT instances, with the latter sustained for more than five hours. 
\end{abstract}

\setboolean{displaycopyright}{true} % Do not include copyright or licensing information in submission.

\begin{document}

\maketitle

\section{Introduction}

Optical network architectures that deliver high-bandwidth and flexible interconnects closer to the compute unit (CU) package are essential for meeting the growing demands of artificial intelligence (AI) workloads and high-performance computing (HPC) \cite{LightwaveFabrics, kumar2024case}. Directly connecting compute nodes via optical interconnects alleviates the dramatic bandwidth taper that is incurred as communication moves off-chip and provides an enhanced degree of reconfigurability \cite{Sirius, RotorNet} that optimizes network architectures to the current workload \cite{Flexpac}. Silicon photonic (SiPh) based interconnects can meet both of these bandwidth scaling and reconfigurability networking requirements \cite{LRL_1} while also being compatible with advanced packaging techniques that can bring optics to the near-package or co-packaged domains. 

Bandwidth steering through wavelength allocation has been demonstrated using multi-port transceivers~\cite{Flexpac}, which also fosters a need for switch-side wavelength reconfigurability. As a result, spatial-and-wavelength selective switches (SWSS) provide a compelling approach for maintaining reconfigurability across parallel WDM channels, routing individual wavelength channels rather than entire ports as in conventional MZI- and MEMS-based switching technologies~\cite{mems, nonvolatilemems, siph_survey, CHEN_REVIEW}. This flexibility is derived from microring resonator (MRR) switching elements that can be thermally tuned to route wavelength channels throughout the fabric~\cite{George_CLEO, Dilated} and can scale~\cite{DAI_SWSS, 8lambda_ofc}. 

More broadly, switching in both space and wavelength has developed rapidly in recent years. Microring-based designs have progressed from multi-wavelength crossbars and fine-resolution wavelength-selective switches to SWSS building blocks~\cite{mw_crossbars, wss-fine-res, push_pull}. Alternative integrated approaches include silicon photonic wavelength cross-connects with 16×16 ports and 16 wavelength channels~\cite{wdm-cross-connect} and SOA-based wavelength-selective cross-connects targeting data center networks~\cite{crossconnect}. Optical circuit switches (OCSes) have also been deployed at the production scale in machine-learning supercomputers~\cite{tpuv4}. As these devices mature, a key open question is how chip-scale SWSS can be integrated into end-to-end AI systems. 

In this work, SWSS is used as the enabling technology for the RADICAL architecture rather than as the subject of a comprehensive device-level comparison across optical switching technologies. Multiple approaches, including large-scale SiPh switching fabrics, MZI-based switch fabrics, MEMS-based optical switches, and wavelength-selective cross-connects, could potentially support reconfigurable direct-connect fabrics~\cite{suzuki, EO_switch, wafer-bonded, flip-chip_mems}, provided that they expose the routing and control functions required by the system. We use an SWSS because it provides spatial path selection together with wavelength-level routing in the same photonic fabric, enabling RADICAL to evaluate wavelength-selective direct-connect reconfiguration in an end-to-end GPU-server cluster. Deploying SWSS-based interconnects into conventional GPU-server clusters and systems requires efficient control and flattening of the network topology in order to extract the system-level advantages for AI and HPC workloads~\cite{Helios, Topoopt, TCP}.

Conventional large-scale data center architectures are typically organized as Clos/fat-tree, or leaf-spine, topologies that rely on electronic packet switches (EPSes) to aggregate and distribute traffic across multiple stages \cite{fattree, VL2}. This adoption is intentional because they prioritize a flatter network with a high degree of connectivity. EPS-based networks also become too expensive and do not scale as well as data centers expand  \cite{bridge, Sato}. Instead, optical circuit switches have been proposed as a replacement for traditional spine-layer EPSes to provide higher bandwidth and point-to-point communication between CUs \cite{CThru, Helios, Hattink, urata2022apollo, optical_sws}. Their integration also enables greater network reconfigurability where the system can dynamically allocate bandwidth to relieve any communication bottlenecks that may arise.

\begin{figure}[!t]
\centering

\begin{minipage}[c]{0.56\linewidth}
    \raggedright
    {\fontfamily{phv}\selectfont\textbf{(a)}}\\[0.2em]
    \centering
    \includegraphics[width=\linewidth]{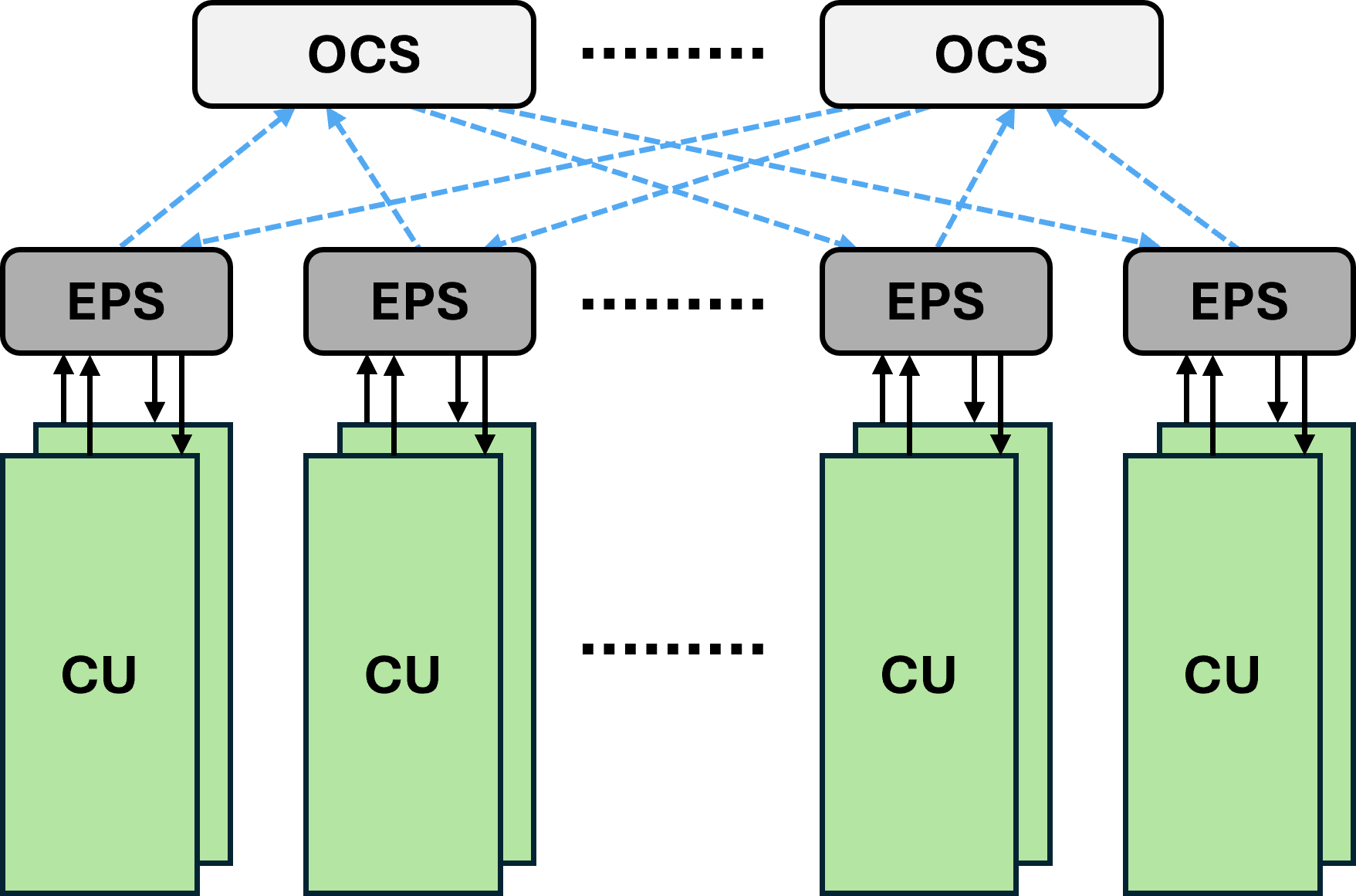}
\end{minipage}
\hfill
\begin{minipage}[c]{0.4\linewidth}
    \raggedright
    {\fontfamily{phv}\selectfont\textbf{(b)}}\\[0.2em]
    \centering
    \includegraphics[width=\linewidth]{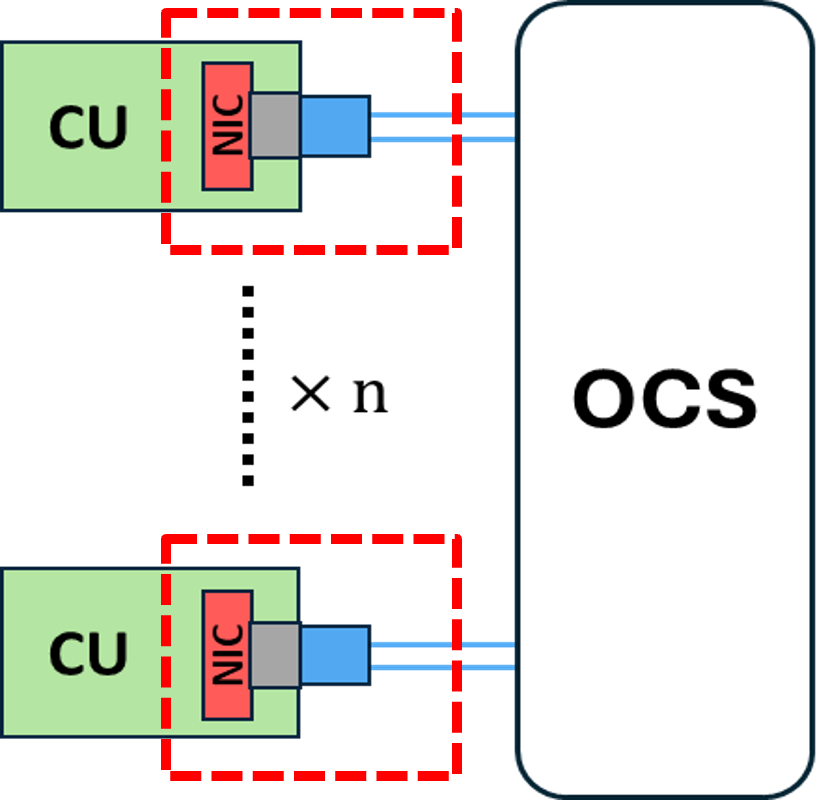}
\end{minipage}
\vspace{0.5em}

\caption{(a) Hybrid switch data center network architecture. (b) Commercial host network interface assumes static connection, which requires modification to be reconfiguration-aware.}
\label{fig:motivation}

\end{figure}

These hybrid-switch networks (Fig. \ref{fig:motivation}(a)) remain constrained by static topologies which limit their ability to exploit the highly deterministic traffic patterns of AI training workloads which can be mapped across the network. This would require a fully reconfigurable network enabled only by directly connecting CUs via a single OCS layer \cite{George_OFC}. Organizing the network in this way has been shown, analytically, to improve workload-completion times over typical leaf-spine hybrid network topologies \cite{Actina}. The system-level advantages have also not been widely shown as a result of the complexity of implementing these direct-connect systems. To do this, the hardware-software interface for reconfigurable networks must be remedied as shown in Fig. \ref{fig:motivation}(b). The commercial interface assumes a static connection that is dropped if a link were to reconfigure.

In this work, we present the silicon-photonic-enabled \underline{R}econfiguration-\underline{A}ware \underline{Di}rect-\underline{C}onnect \underline{A}I C\underline{l}uster (RADICAL), a full-stack demonstration of a reconfigurable optical direct-connect fabric built from an MRR-based $4\times4\times4\lambda$ SWSS, commercial GPU servers, standard optical transceivers, and a reconfiguration-aware Linux host interface. A central system challenge for reconfigurable networks is how conventional host networking interfaces assume a static link connection which is dropped when the physical connectivity changes. RADICAL addresses this problem by coordinating SWSS control with a Linux packet scheduler that tracks the active destination and buffers traffic during reconfiguration phases. This integration enables bidirectional 25 Gbps SWSS-routed links, sustained bulk data transfer, and multitenant AI training across dynamically reconfigured link topologies. This paper is organized as follows: Section 2 describes the SWSS architecture, Section 3 characterizes the packaged $4\times4\times4\lambda$ switch, Section 4 presents RADICAL's direct-connect architecture, Section 5 describes the hardware and software integration, Section 6 evaluates the system experimentally, Section 7 discusses scalability and future directions, and Section 8 concludes the paper.

\section{SWSS Architecture}
\label{sec:swss}

Spatial-and-wavelength selective switching enables two levels of routing capability for WDM-based communication systems. To implement SWSS-based interconnects, the SWSS architecture must provide high bandwidth, flexible wavelength routing, and an established control scheme. These three specifications are required for using this device in a data center network environment.

\subsection{Switching Elements and Fabric}

\begin{figure}[h]
    \centering
    \includegraphics[width=0.95\columnwidth]{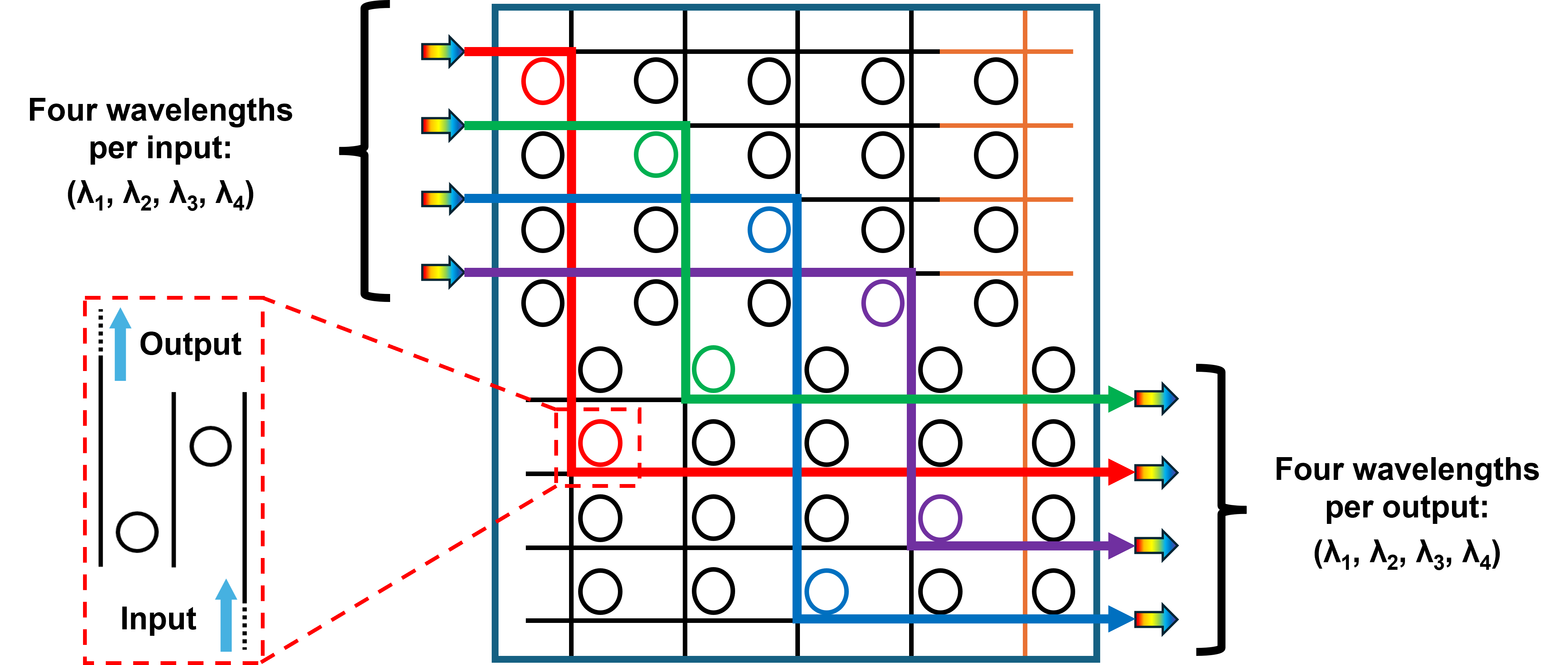}
    \caption{$4\times 4\times 4\lambda$ SWSS Architecture that consists of four input ports, four output ports, and four wavelengths per port. Each switching element in the fabric contains two thermally tuned MRRs as shown by the highlighted section in red.}
    \label{fig:swss_arch}
\end{figure}

These switching fabrics can route parallel wavelength channels as well as route per spatial port to provide connectivity across the network. These fabrics have two forms of tunability: (1) coarse-grained tunability across spatial ports, (2) fine-grained tunability across wavelength channels. Because of the two forms of switching, SWSS-fabrics comprise $n$ input ports, $n$ output ports, and $\lambda$ wavelength channels per input port. Thus, the switch matrix can be described by the following: $n\times n \times\lambda$. Fig.~\ref{fig:swss_arch} shows a $4\times 4\times 4\lambda$ SWSS architecture which will be used throughout this work.

The switching elements used in the SWSS are made up of two thermally-tuned MRR filters. As shown in Fig.~\ref{fig:swss_arch}, the two filters are arranged such that the light from the input buses can be dropped throughout the rest of the fabric. The light is first filtered from the input bus and dropped into an intermediary waveguide that is coupled to the second filter. Then, the second filter is tuned to the same wavelength, coupling the selected signal into the fabric via the vertical buses and towards the output buses. The fabric contains five vertical buses: four primary buses and a fifth bus which provides additional routing flexibility for the switch. Having two MRRs for each switching element allows for increased crosstalk suppression between spatial ports \cite{DAI_SWSS}. It is here that these switching elements are utilized to route individual WDM signals throughout the switch. 

The SWSS switching elements are arranged into a double-crosspoint topology which consists of two halves where signals are added and then dropped. The first half of the SWSS fabric is a de-multiplexing stage with wavelength selectivity. Each WDM channel is filtered from the input bus and dropped to a vertical bus (Fig.~\ref{fig:swss_arch}). Then, the second half of the fabric is a wavelength selective multiplexing stage which adds each of the dropped WDM channels to the correct horizontal output bus to be sent off the chip. The fabric is wrapped around itself such that the vertical buses are connected at both ends to provide full connectivity across the fabric. 

An example of routing paths throughout the switch is shown in Fig.~\ref{fig:swss_arch}. A practical application would connect each spatial input port to a transmitter, and each output port to a receiver. For a single WDM signal to be fully routed, two SWSS switching elements (four MRRs) must have their resonances aligned and tuned to that signal. Controlling each of these resonances is a crucial part of the RADICAL hardware integration.

\subsection{Device Design and Control}

\begin{figure}[t]
\centering

\begin{minipage}[c]{0.48\linewidth}
    \raggedright
    {\fontfamily{phv}\selectfont\textbf{(a)}}\\[0.2em]
    \centering
    \includegraphics[width=\linewidth]{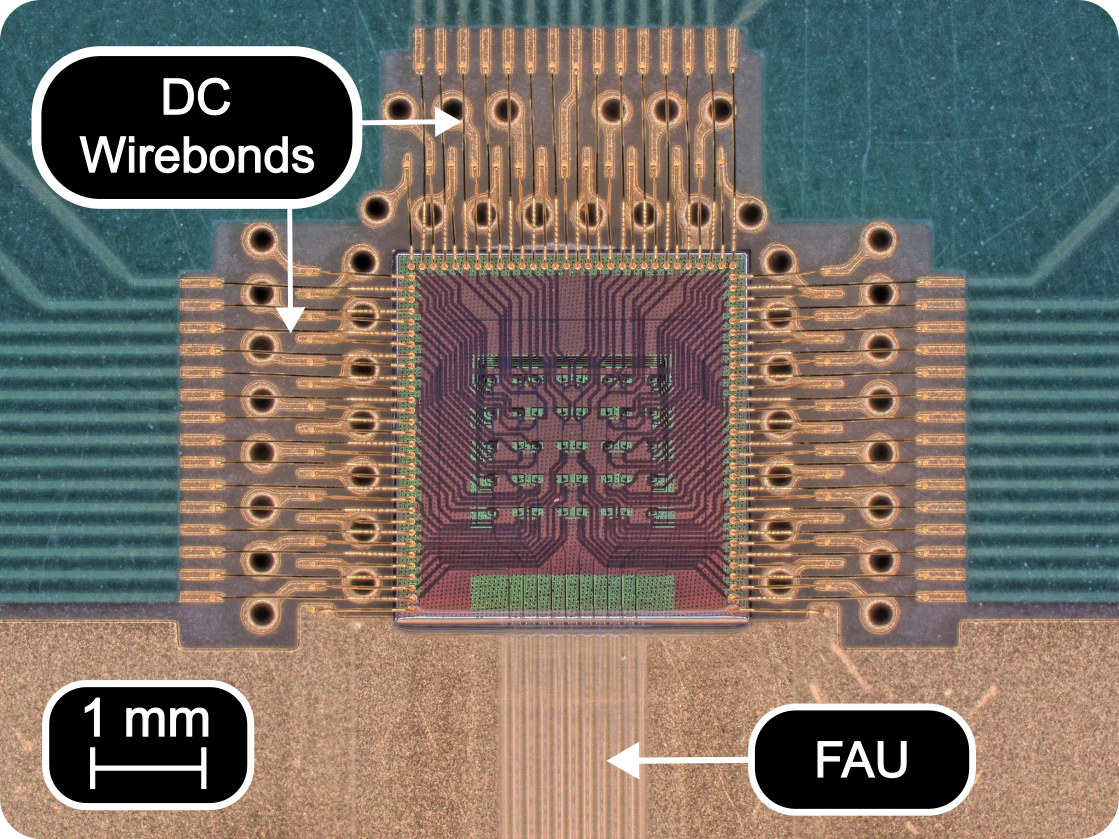}
\end{minipage}
\hfill
\begin{minipage}[c]{0.48\linewidth}
    \raggedright
    {\fontfamily{phv}\selectfont\textbf{(b)}}\\[0.2em]
    \centering
    \includegraphics[width=\linewidth]{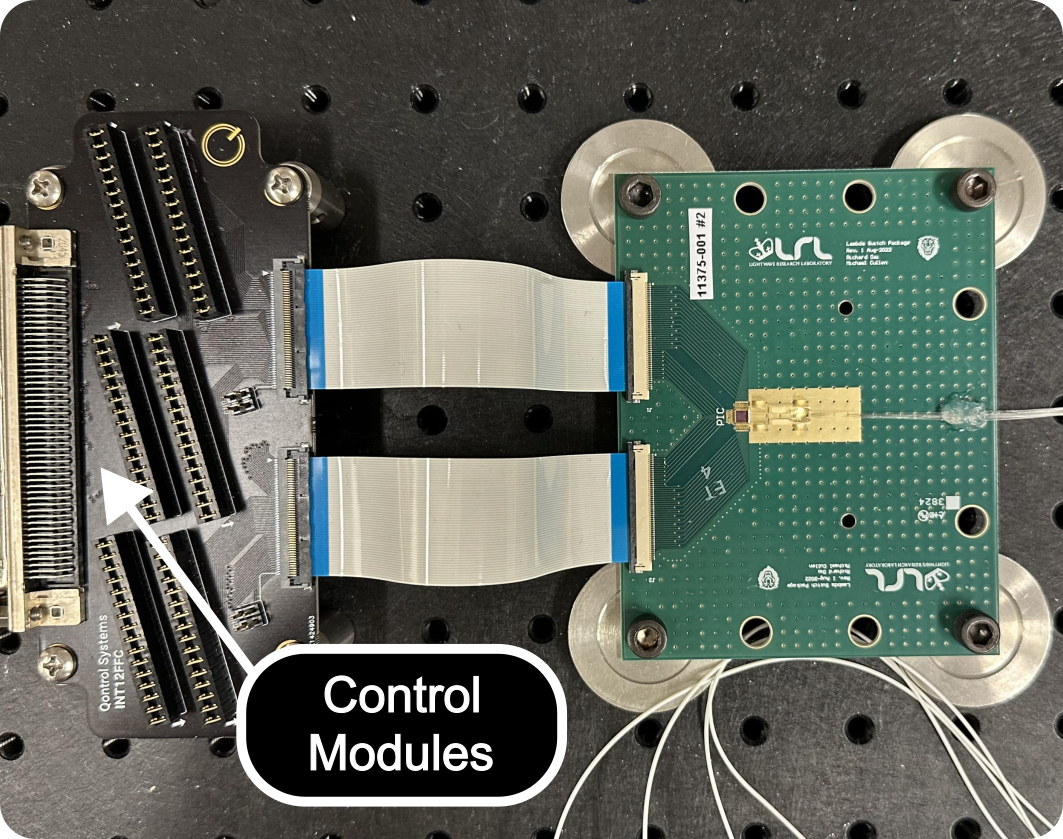}
\end{minipage}
\vspace{0.5em}

\caption{(a) Micrograph image of the $4\times4\times4\lambda$ switch with DC wirebonds to a PCB. (b) Fully packaged switch setup with PIC and FAU mounted on a PCB which interfaces with the switch control modules.}
\label{fig:pic}

\end{figure}

We designed a $4\times4\times4\lambda$ SWSS for this work and for use in RADICAL. The photonic integrated circuit (PIC) was designed by Columbia and fabricated by AIM Photonics. It consists of 80 individual MRR filters that can be thermally-tuned to match the desired resonance. Light is coupled into the device via edge couplers that are spaced at 127 $\mu$m apart. The PIC itself has a total of 40 switching elements or 80 MRRs. Each filter has a free spectral range (FSR) of $~25\,$nm, which enables increased flexibility for choosing the spacing and wavelengths of our input WDM signals. The resonance of each filter is thermally-tuned by setting a DC control voltage to respective MRR. The control signals are routed to metal pads that run along the perimeter of the device as shown in Fig.~\ref{fig:pic}(a). A PCB was designed to mount the PIC, the fiber array unit (FAU), and provide wirebond pads corresponding to each electrical pad on the PIC. The optical packaging, alignment, and electrical wirebonding were performed by Optelligent in Costa Mesa, CA. 

We control the $4\times4\times4\lambda$ switch by first determining the tuning efficiency of the MRR switching elements and then applying DC control signals to the device. The switch PCB transfers each of the signals to ribbon cable connectors that interface directly with the control modules that can be configured via Python. Fig.~\ref{fig:pic}(b) shows the fully packaged $4\times4\times4\lambda$ switch, and our control interface. The control modules used in our setup are 12x Qontrol Q8iv programmable drivers. By setting DC voltages to the thermal heaters of the filters, the resonance can be shifted to match that of the WDM signal being routed on that path.

Thus, to integrate a SWSS device, the control loop must thermally tune each of these MRRs to establish the desired route for that period of time. When the switch is fully loaded with signals, four WDM signals per port, an automated routing algorithm can be used to tune the entire switch.

\subsection{Automated Routing Algorithm}
\label{sec:algo}

The automated routing algorithm as presented at CLEO '25 \cite{George_CLEO} and described in Alg.~\ref{alg:route} for the SWSS starts with an input demand matrix for each input signal. Each WDM signal has an input $I$, output $O$, and wavelength $\lambda$ that is used to construct the tuple: $(I, O, \lambda$). We then construct a graph $G(V,E)$ to represent the SWSS fabric where $v \in V$ is a distinct signal, and each edge $e \in E$ represents an association that two signal tuples have by sharing the same $I$, $O$, and/or $\lambda$. A vertex-coloring algorithm is then applied to the SWSS graph where each vertex is assigned a color that does not match any adjacent vertices. Because the SWSS topology is fixed and evaluated routing demands are admissible, the coloring step is used as a resource-assignment procedure rather than as a general minimum-coloring problem, which is NP-hard. These colors are then mapped to one of the five vertical buses on the switch fabric. With each path mapped, the signals can be assigned MRR channels, and a pre-calibrated tuning voltage function can be called. This function uses the tuning efficiency of the MRR heaters to determine the voltage required to match the filter resonance to the wavelength of the specific signal.

\begin{algorithm}[!t]
\caption{Automated SWSS Routing Algorithm}\label{alg:route}
\begin{algorithmic}[1]

\Function{swssGraph}{$T$}
    \For{each tuple $\left(I_i, O_i, \lambda_i\right) \in T$}
        \State{Add vertex $v_i$ to $V$}
    \EndFor
    \For{each pair of vertices $(v_i, v_j)$}
        \If{$I_i = I_j~\text{or}~O_i = O_j~\text{or}~\lambda_i = \lambda_j$}
            \State{Add edge $e_k$ to $E$}
        \EndIf
    \EndFor
    \State \textbf{return}{ $G(V,E)$}
\EndFunction
\Statex
\Procedure{routeSWSS}{$D$} \Comment{$D$: input demand matrix}
    \State $T \gets$ signal tuples $\{(I_i, O_i, \lambda_i)\}$ from $D$
    \State $G(V, E) \gets \Call{swssGraph}{T}$
    \State $G_c(V,E) \gets \Call{vertexColoring}{G}$
    \State Assign distinct vertex colors to vertical drop buses
    \For{$v_i \in G_c$}
        \State{Map $(I_i, O_i, \lambda_i)$ to SWSS MRR channels}
        \State{Compute heater voltages from pre-calibration}
        \State{Apply voltages to align MRR resonances to $\lambda_i$}
    \EndFor
    \label{swss_routing}
\EndProcedure

\end{algorithmic}
\end{algorithm}

\section{SWSS Characterization}
\label{sec:char}

\begin{figure*}[ht]
\centering
\includegraphics[width=\textwidth]{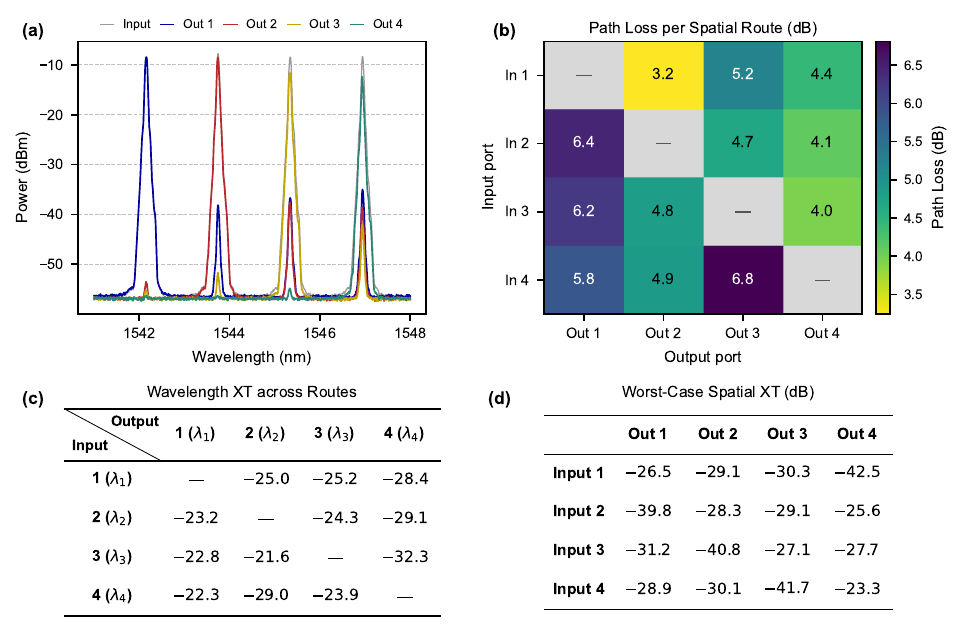}
\caption{(a) Multi-wavelength routing spectra showing the original combined input and the WDM signal peaks after routing across the switch. (b) Heatmap of the path loss associated with the spatial switch paths that excludes self-routes since no switching is required. (c) Wavelength crosstalk measurements across all wavelength channels. (d) Worst-case spatial crosstalk measurements across all spatial ports.}
\label{fig:device_chars}
\end{figure*}

The following section characterizes the performance of the $4\times4\times4\lambda$ SWSS by discussing multi-wavelength routing spectra, path loss across the switch, wavelength crosstalk (XT$_\lambda$), and spatial crosstalk (XT$_s$). We used these specifications to determine how the SWSS would be used in the RADICAL platform. For the SWSS characterization tests, we used the following four wavelengths because they were the transceiver wavelengths in RADICAL and standard ITU channels. These are shown in Table~\ref{tab:wavelengths}. These wavelengths have a uniform $1.6$ nm (200~GHz) spacing on the ITU grid. The following device-level and system-level experiments will repeatedly use these four wavelength channels. 

\begin{table}[htbp]
\centering
\caption{\bf RADICAL Transceiver Wavelength Channels}
\begin{tabular}{ccccc}
\hline
 & $\lambda_1$ & $\lambda_2$ & $\lambda_3$ & $\lambda_4$ \\
\hline
Wavelength (nm) & $1542.14$ & $1543.73$ & $1545.32$ & $1546.92$ \\
ITU Channel & C44 & C42 & C40 & C38 \\
\hline
\end{tabular}
  \label{tab:wavelengths}
\end{table}

\subsection{Multi-wavelength Routing Spectra}

We captured a multi-wavelength routing spectra by combining the four wavelength channels from Table $\ref{tab:wavelengths}$ using an optical combiner. All four wavelengths were then sent to a single spatial input and routed throughout the switch. To show full tunability across spatial ports and wavelength channels, each WDM signal was sent to a different output.

Fig. \ref{fig:device_chars}(a) shows the input and routed spectra of the four wavelength channels. This plot compares the peak power of an unrouted and routed wavelength channel. The input spectrum illustrates the peak power of each channel before these signals were routed through the switch. These are measured at the throughport of output 1. Then, we simultaneously routed each wavelength to a different spatial port. As shown, $\lambda_1$ is sent to output 1, $\lambda_2$ is sent to output 2, $\lambda_3$ is sent to output 3, and $\lambda_4$ is sent to output 4. The resulting four spectra show each signal at the output after tuning the four MRRs per path. For each wavelength channel, the routed power and input power can be compared to determine how well the MRR switching elements are tuned to that channel.

\subsection{Path Loss Across Spatial Ports}

The path loss is defined as the power loss associated with routing a single WDM signal from input to output. It compares the following two power measurements:
\begin{equation}
    \text{Path Loss (dB)} = P_\text{loopback} - P_{i,j},
\end{equation}
where $(i,j)$ denote I/O and are measured at the same wavelength. This is comparing the power measured at the output of a loopback waveguide to the power after routing a signal to various output ports. 

The measurements were made by first inputting a $+6$ dBm signal from the laser source into the loopback path and measuring the output power. We then repeated this for all four wavelengths. Then, we assigned a single wavelength to each input port (input 1 used $\lambda_1$ etc.), and routed each signal to the remaining outputs, excluding self-routes since they do not require MRR tuning. We repeated this power measurement until all paths were measured. Fig.~\ref{fig:device_chars}(b) shows the results of the path loss measurements where redundant measurements were averaged. The results show that the path loss ranges from $3.2$ dB to $6.8$ dB. These differences are caused by both variations in the path (more crossings, throughports) and resonance mismatch between all four MRRs. These measurements were used to establish the optical margin of the SWSS integrated into RADICAL and to verify that every calibrated route remained within a practical optical power budget. As shown in Section~\ref{sec:eval}, RADICAL sustained data transfer and AI training workloads across this full range of path loss, supporting the feasibility of using the SWSS for multi-wavelength reconfigurable interconnects.

\subsection{Wavelength Crosstalk}
\label{sec:wl}

Wavelength crosstalk ($XT_\lambda$) is defined as the difference between the worst-case leakage power of an undesired WDM signal and the power of the desired WDM signal. As an equation, it can be defined as:
\begin{equation}
    XT\,\text{(dB)} = P_\text{leak} - P_\text{signal}.
    \label{eqn:xt}
\end{equation}
Using the same testing arrangement as in the path loss experiment, each input is assigned a wavelength, and we route these wavelengths across all spatial ports, excluding self-routes. 

The results shown in the table in Fig.~\ref{fig:device_chars} illustrate that across the entire switch, we have $>20$~dB of crosstalk suppression across wavelength channels, accounting for the worst-case. On average, the switch paths have $|XT_\lambda| > 25\,$~dB. The switch exhibits low crosstalk, which is essential in the operation of RADICAL. These results indicate effective wavelength isolation across measured switching paths. 

\subsection{Spatial Crosstalk}
\label{sec:sp}

Spatial crosstalk ($XT_s$) is defined as the isolation between WDM signals that are launched simultaneously into the same input port and individually routed to different output ports. It is measured using the same setup as the multi-wavelength routing spectra. Four $+6$~dBm signals are multiplexed into one input port, and $\lambda_1$ through $\lambda_4$ are routed to output ports 1 through 4, respectively. We then compare the power of the leakage signals to the desired signal power as in Eq.~\ref{eqn:xt}.

Fig.~\ref{fig:device_chars}(d) shows a table of worst-case $XT_s$ measurements. Worst-case is defined as the highest leakage power peak from the undesired signals. Again, the SWSS shows good isolation across all spatial ports where we observe $|XT_s| > 23$~dB, and on average have $>30$ dB of suppression across all spatial channels. In both crosstalk cases, we find that the SWSS provides good isolation between adjacent WDM signals. 

\subsection{Signal Integrity}

To verify the signal integrity under system-relevant operating conditions, we measured the bit error rate (BER) of the switch paths at 25 Gbps, matching the data rate of the SFP28 transceivers used in RADICAL. Before transmitting real data, we tested whether a signal could be received error-free (BER$\,<10^{-12}$) across SWSS I/O paths. We selected four representative routes: Input 1 to Output 3 at $1542.14$ nm, Input 2 to Output 1 at $1543.73$ nm, Input 3 to Output 2 at $1545.32$ nm, and Input 4 to Output 3 at $1546.92$ nm. Three of these routes had losses above the 5.0 dB mean shown in Fig. \ref{fig:device_chars}(b), while the Input 3 to Output 2 route had a loss of $4.8$ dB. The measurement input was a non-return-to-zero (NRZ) PRBS $2^{31}-1$ signal modulated at $25$ Gbps.

\begin{figure}[ht]
\centering

\begin{minipage}[t]{0.62\linewidth}
    \raggedright
    {\fontfamily{phv}\selectfont\textbf{(a)}}\\[0.2em]
    \centering
    \includegraphics[width=\linewidth]{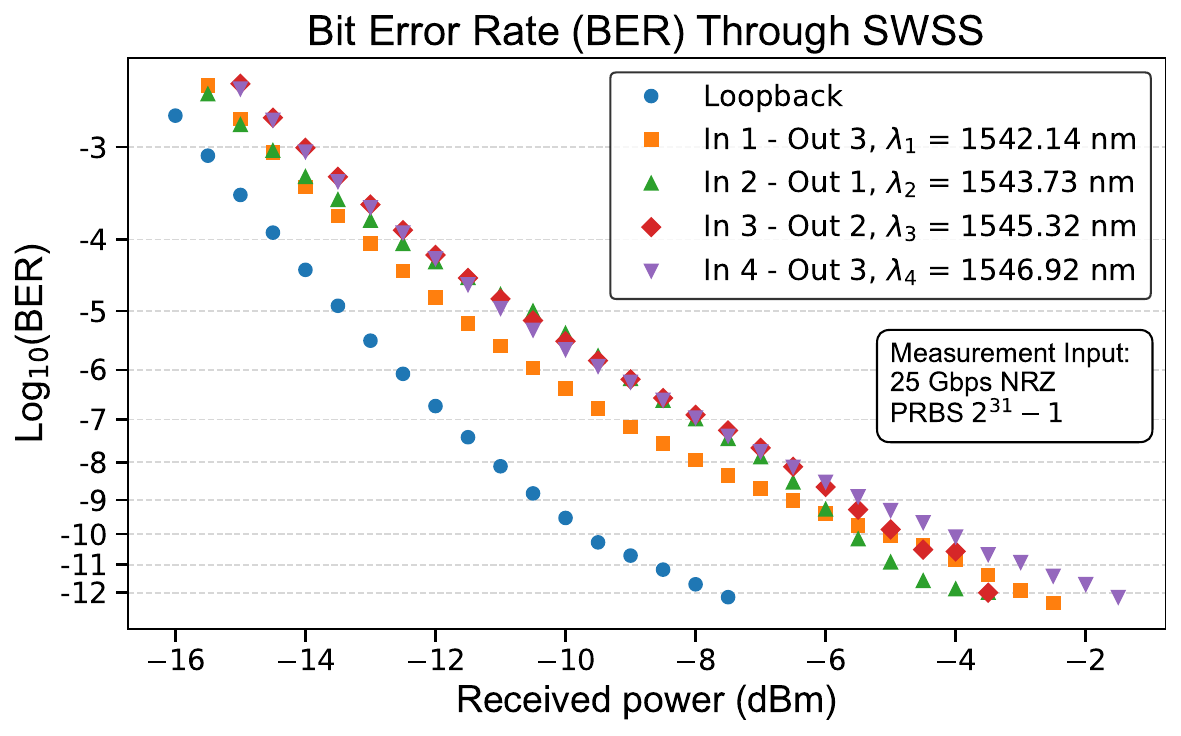}
\end{minipage}
\hfill
\begin{minipage}[t]{0.34\linewidth}
    \raggedright
    {\fontfamily{phv}\selectfont\textbf{(b)}}\\[0.2em]
    \centering
    \includegraphics[width=\linewidth]{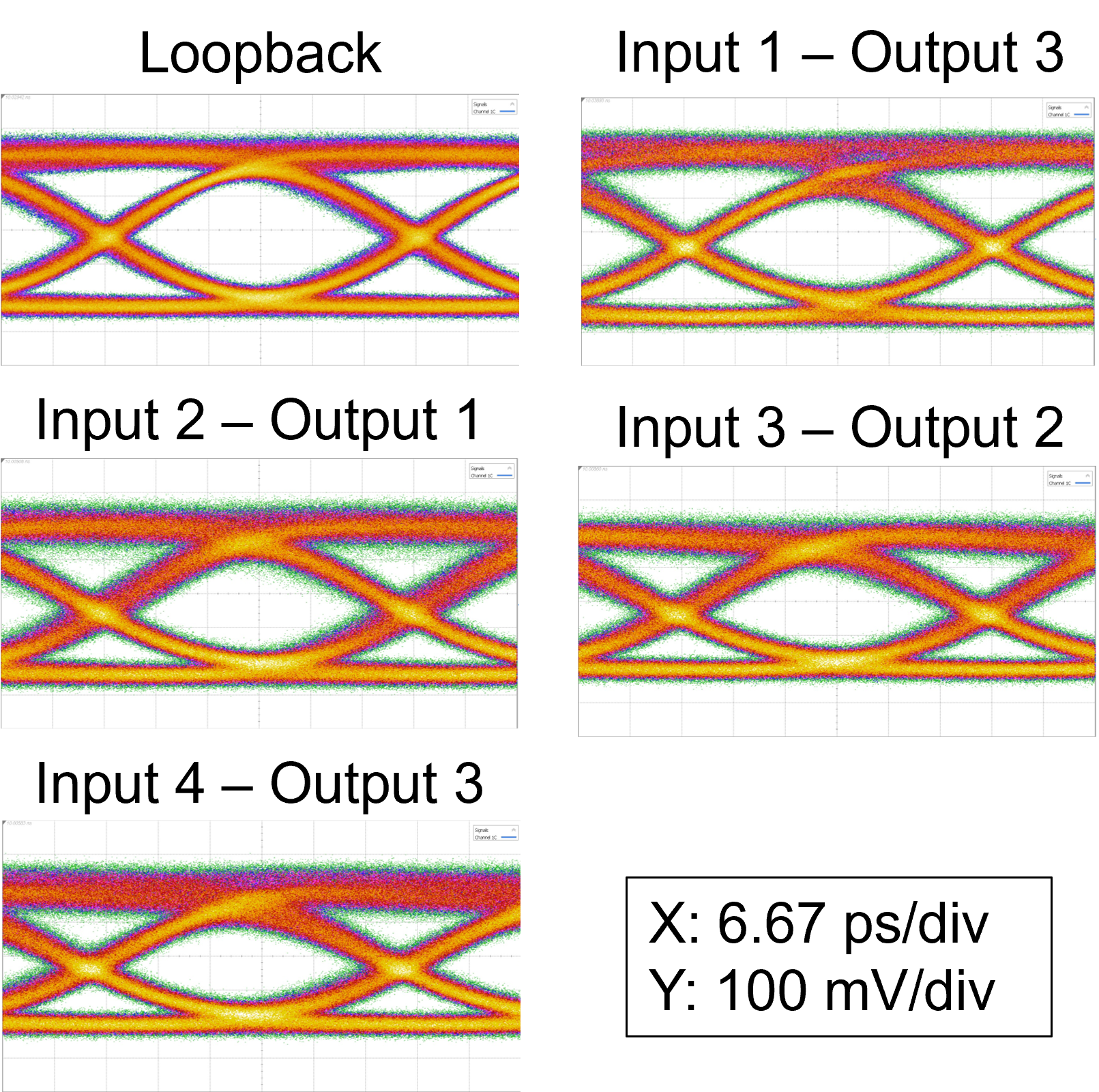}
\end{minipage}
\vspace{0.5em}

\caption{(a) Bit Error Rate (BER) through various I/O routes and wavelengths; each signal begins at an error-free BER. (b) Eye diagrams for all four switch routes and the loopback at $25$ Gbps.}
\label{fig:ber}
\end{figure}

From Fig.~\ref{fig:ber}(a), we can derive the power penalty of each path as the difference in the error-free received power of the loopback versus the routed paths. The power penalty ranges from $4$ dB to $6$ dB with an average power penalty of $4.75$ dB across the four selected I/O routes. Even with this higher penalty, our results in Section 6 show that we overcome it in our hardware integration.

\section{Direct-Connect Architecture}

Direct-connect network architectures have been proposed as a way to bring a higher bandwidth networking layer closer to the compute node \cite{LightwaveFabrics, Topoopt}. This is achieved by condensing the number of switching layers in the overall network. The Helios architecture \cite{Helios} was one of the earliest to propose this by implementing a hybrid switch topology. In that architecture, electrical pod switches aggregate the servers within each pod, and the core layer contains both EPSes and OCSes that carry different traffic between pods. With this, the architecture introduced limited flexibility to dynamically allocate its bandwidth or links across the network. However, because only a subset of the network is optically reconfigurable, hybrid EPS/OCS designs still constrain bandwidth steering through the remaining fixed electronic switching stages and static topology. Direct-connect optical fabrics therefore provide a different design point, where the reconfigurable switching layer can be exposed uniformly across the communicating compute nodes. 

\begin{figure}[!h]
    \centering
    \includegraphics[width=0.7\linewidth]{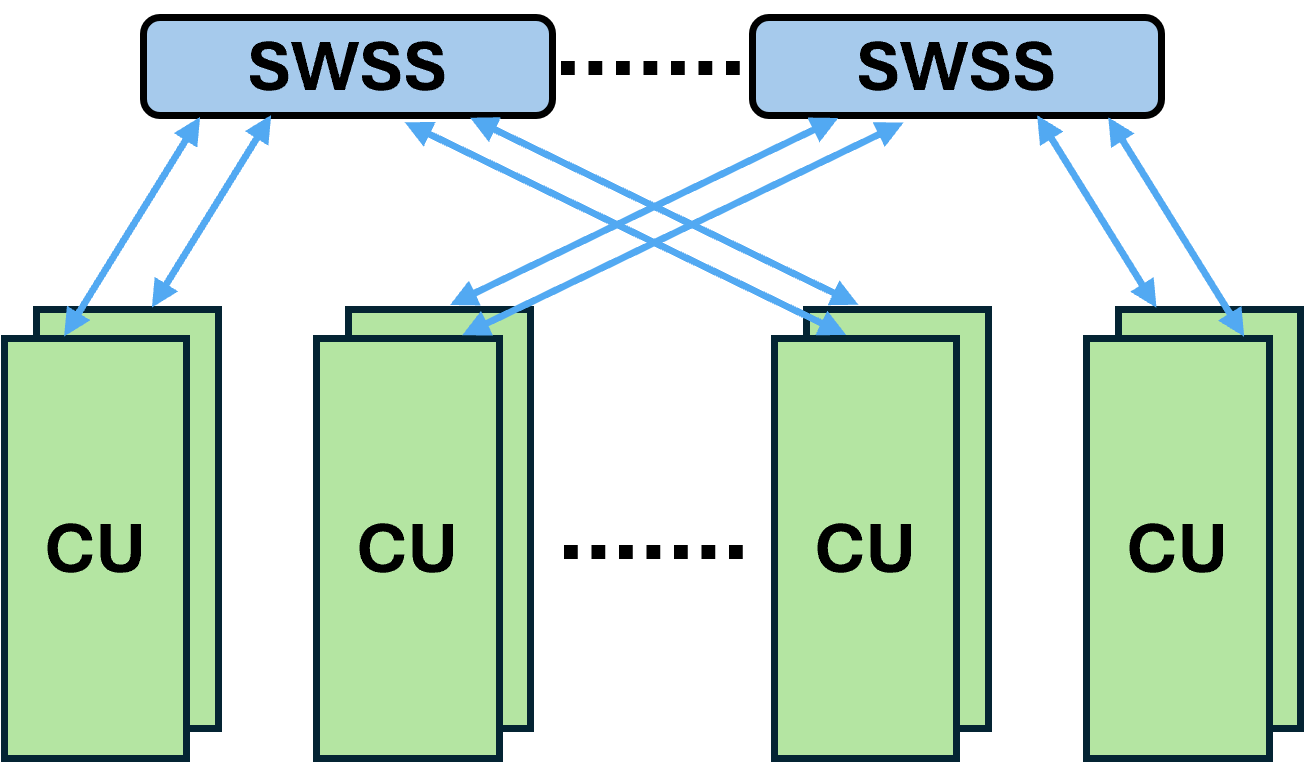}
    \caption{RADICAL's direct-connect architecture with a single SWSS interconnect layer.}
    \vspace{-0.5em}
    \label{fig:radical}
\end{figure}

\begin{figure*}[!t]
\centering

\begin{minipage}[c]{\linewidth}
    \raggedright
    % {\fontfamily{phv}\selectfont\textbf{(a)}}\\[0.2em]
    \centering
    \includegraphics[width=\linewidth]{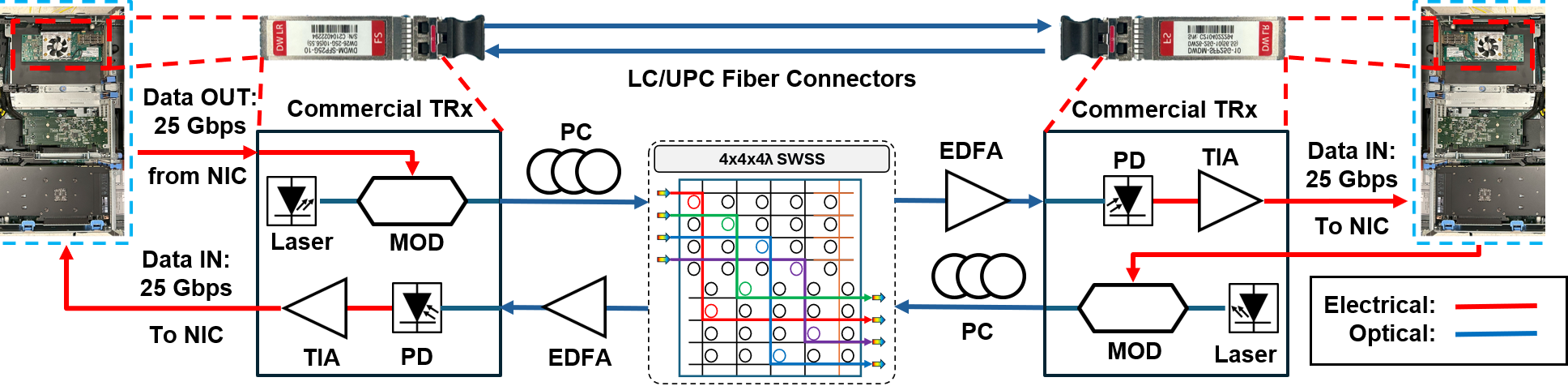}
\end{minipage}

\caption{End-to-end bidirectional SWSS link diagram which uses commercial SFP28 transceivers and interfaces directly to our custom $4\times4\times4\lambda$ switch.}
\label{fig:full_link}

\end{figure*}

Prior efforts such as RotorNet \cite{RotorNet}, Sirius \cite{Sirius}, and earlier Columbia work \cite{Hattink} have demonstrated the potential of reconfigurable direct-connect networks, but their deployment assumptions differ substantially from RADICAL. RotorNet achieves direct-connect optical networking through specialized switching and host-side assumptions with predetermined routing states. Sirius explores a flat optical switching layer based on nanosecond-scale switching technology, which is attractive for highly dynamic traffic but is not always required for deterministic AI training workloads \cite{Actina}. Previous work at Columbia developed a software-defined control plane that combines optical reconfigurability with electronic packet switching. Google's Jupiter \cite{Jupiter} demonstrates the value of at-scale optical circuit switching in production data center fabrics, but at the block-interconnect level rather than as an SWSS-enabled direct-connect fabric between GPU servers. In contrast, RADICAL demonstrates a wavelength-selective SiPh switching fabric integrated with commercial GPU servers and transceivers, a reconfiguration-aware Linux kernel, and real AI training workloads. 

In this work, RADICAL implements a reconfiguration-aware direct-connect network through a full hardware and software integration, Fig. \ref{fig:radical}. We implement SWSS-based interconnects that can dynamically allocate links across a cluster of GPU-servers. Our software integration consists of implementing a reconfiguration-aware Linux packet scheduler that can notify and track the current links established and jobs associated with this topology. Using both of these elements, the network can be optimized to track the traffic patterns of the AI workload being trained at that point in time. 

The graph-coloring procedure used by RADICAL configures the SWSS switch state for a requested set of spatial-wavelength connections. In our implementation, each vertex corresponds to a requested input-output-wavelength tuple, so the graph size is set by the per-switch configuration rather than by the total number of computing units in a larger fabric. For higher-radix SWSS designs, such as radix 8 or radix 16, workload-aware switch states can be computed offline and stored as calibrated lookup-table entries before reconfiguration. The RADICAL system can be viewed as a hardware implementation of the direct-connect optical fabric proposed by OCSBCube in a single network dimension \cite{Actina}. Scaling this design to larger fabrics would follow the same dimensional composition approach, where multiple smaller-radix optical switches are arranged across network dimensions. For example, a radix-16 SWSS could support $16^2 = 256$ chips in a two-dimensional fabric and $16^3 = 4,096$ chips in a three-dimensional fabric. In a scaled system, reconfiguration would be coordinated by selecting the precomputed states for the relevant SWSS devices at workload or training-phase boundaries. In future scaled-up systems, device-integration work would need to address higher-radix SWSS packaging, denser wavelength planning, and calibration methods that preserve optical margin as more rings and wavelength channels are integrated. 
The next section of this work will cover the RADICAL hardware and software integration and the experiments we ran to evaluate the system. 

\section{Hardware-Software Integration (RADICAL)}

We integrated our custom $4\times4\times4\lambda$ switch within a commercial compute and transceiver environment to build RADICAL. We solved the host network interface for reconfigurable networks by modifying the Linux packet scheduler.

\subsection{Hardware Integration}

\begin{figure}[!h]
    \centering
    \includegraphics[width=0.98\columnwidth]{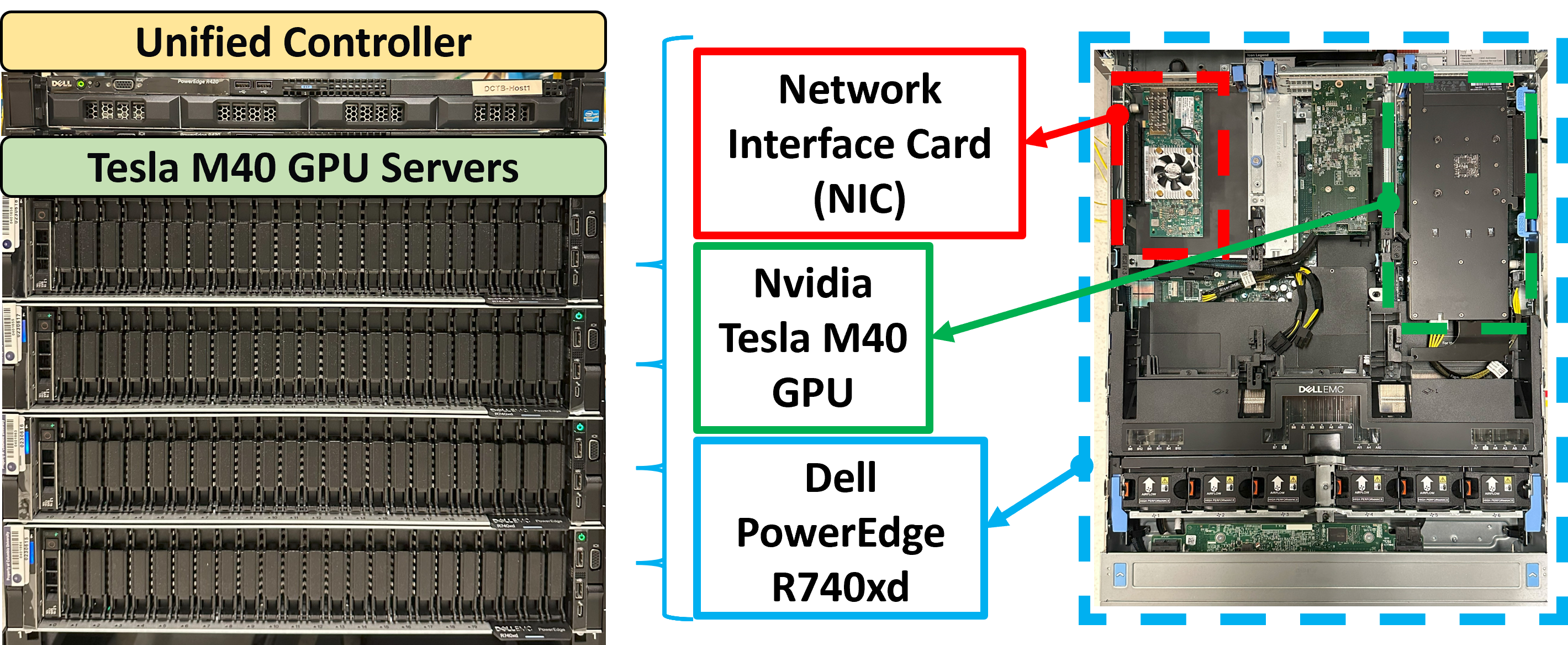}
    \caption{RADICAL compute hardware, including NVIDIA Tesla M40 GPUs and commodity NICs. The unified controller server communicates with the GPU-servers via a top-of-rack (ToR) switch.}
    \label{fig:GPUservers}
\end{figure}

The hardware integration begins with four compute units that will be executing our AI workloads. In Fig. \ref{fig:GPUservers}, we use four Dell PowerEdge R740xd servers, each of which has an NVIDIA Tesla M40 GPU and an HPE 620QSFP28 NIC. We use four SFP28 transceivers that transmit at one of the four wavelengths defined in Table \ref{tab:wavelengths}. We interface the QSFP ports with SFP28 transceivers by using QSFP to SFP form-factor converters. A unified controller server is used to relay ICMP network status messages to the hosts (GPU-servers) during transmit and reconfiguration network phases.

An end-to-end bidirectional link diagram routed through the SWSS is shown in Fig. \ref{fig:full_link}. Four simultaneous and bidirectional links are connected to and routed through the switch at any given time. The hosts generate traffic that is sent to the NIC to be transmitted to the network over fiber at $25$ Gbps. The SWSS is tuned by setting the control voltages to the MRR heaters to establish routes through the switch. To compensate for fiber-to-chip coupling loss, we use an EDFA at each SWSS output to ensure the WDM signals stay within the operating range of the SFP28 receiver. The 25 Gbps line rate is set by the C-band SFP28 host transceivers, which are readily available and align with the wavelength range of the SWSS. Moreover, RADICAL targets a “slow and wide” approach to bandwidth scaling, in which aggregate bandwidth comes from many parallel wavelength channels at moderate per-wavelength rates rather than from very high-speed optics. For instance, a WDM host interface with four 25 Gbps wavelength channels per port would provide 100 Gbps of aggregate per-port bandwidth before considering link budget and implementation overheads. 

The SWSS, as described previously, is tuned using 80 voltage  control signals from the Qontrol Q8iv control modules. These signals are set in software by the unified controller server, which updates the filter voltages depending on the routes that need to be established for a given link topology. During reconfiguration phases, these filters must be reset, and the new paths need to be tuned. During reconfiguration, the rings associated with the previous paths are turned off, and the rings associated with the new paths are tuned to their calibrated settings. This control software works in tandem with the host's reconfiguration-aware Linux kernel to ensure proper operation of the system.

\begin{figure}[h!]
    \centering
    \includegraphics[width=0.98\columnwidth]{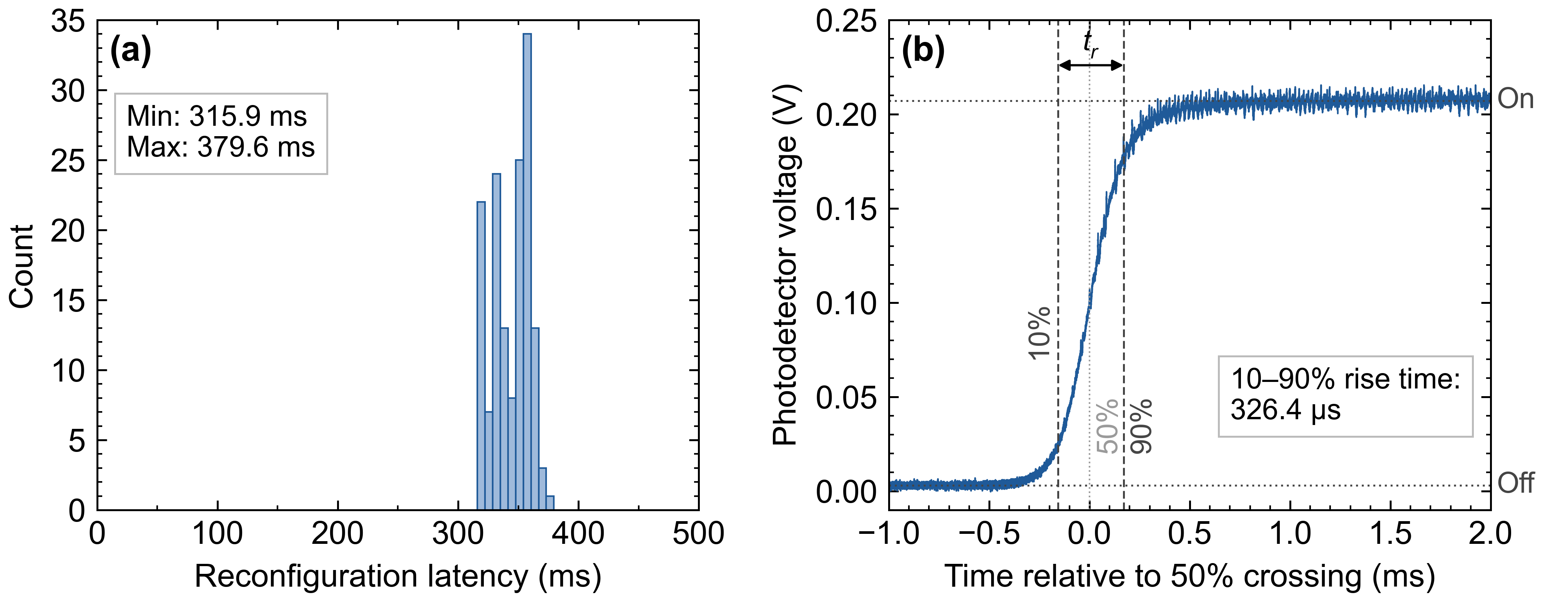}
    \caption{(a) Link reconfiguration latency distribution measured from hosts. (b) Representative Input 1 to Output 2 optical transition. The dashed vertical lines indicate 10\%, 50\%, and 90\% crossing points used for the rise-time calculation.}
    \label{fig:latency}
\end{figure}
 
The current system configures the SWSS paths using a lookup table of calibrated control voltages and the graph-coloring control algorithm that covers all 80 rings (Section~\ref{sec:swss}\ref{sec:algo}, Alg.~\ref{alg:route}). Because AI training communication patterns are largely deterministic, the required switch states can be precomputed from the workload and selected at training-phase boundaries or per workload rather than at burst-level timescales~\cite{Topoopt} The crosstalk measurements in Sections~\ref{sec:char}\ref{sec:wl}-\ref{sec:sp} characterize the spatial and wavelength isolation for the calibrated wavelength grid and path configurations used in RADICAL. If future designs use more closely spaced wavelengths or higher-radix SWSS designs, these calibrated voltages would need to be updated to account for changed resonance spacing and path loss. In this approach, increasing the number of rings mainly adds to the offline calibration effort, while reconfiguration itself consists of applying stored switch states. 

We also measured the end-to-end reconfiguration time observed by the hosts. Across 150 consecutive network reconfigurations, the measured latency ranges from $315.9$ ms to $379.6$ ms, as shown in Fig. \ref{fig:latency}(a). This value should be interpreted as a prototype-level system-control latency rather than an intrinsic limitation of the SWSS fabric. To separate the device-level optical transition from this end-to-end system measurement, we measured the rise time for a representative SWSS path from Input 1 to Output 2. Across five runs, the measured 10-90\% optical rise time was $333.5 \pm 11.2 ~\mu$s, and a representative optical transition is shown in Fig. \ref{fig:latency}(b). This sub-millisecond transition is orders of magnitude shorter than the measured end-to-end reconfiguration latency. In this integrated prototype, the latter includes transceiver relocking, host notification, and serial communication with the external Qontrol controller modules. Taken together, Fig. \ref{fig:latency} shows that the optical switching event is not the limiting factor in the current implementation. Rather, the end-to-end latency points to the surrounding NIC, transceiver, and system control stack as the primary targets for latency optimization. For AI training workloads, this overhead can be amortized by using coarse-grained, workload-aware reconfiguration strategies \cite{tpuv4, Jupiter} However, for more frequent topology reconfigurations, microsecond-scale end-to-end latency is required to see appreciable performance gains \cite{Actina}.

\subsection{Software Integration}

\begin{figure}[!h]
    \centering
    \includegraphics[width=\columnwidth]{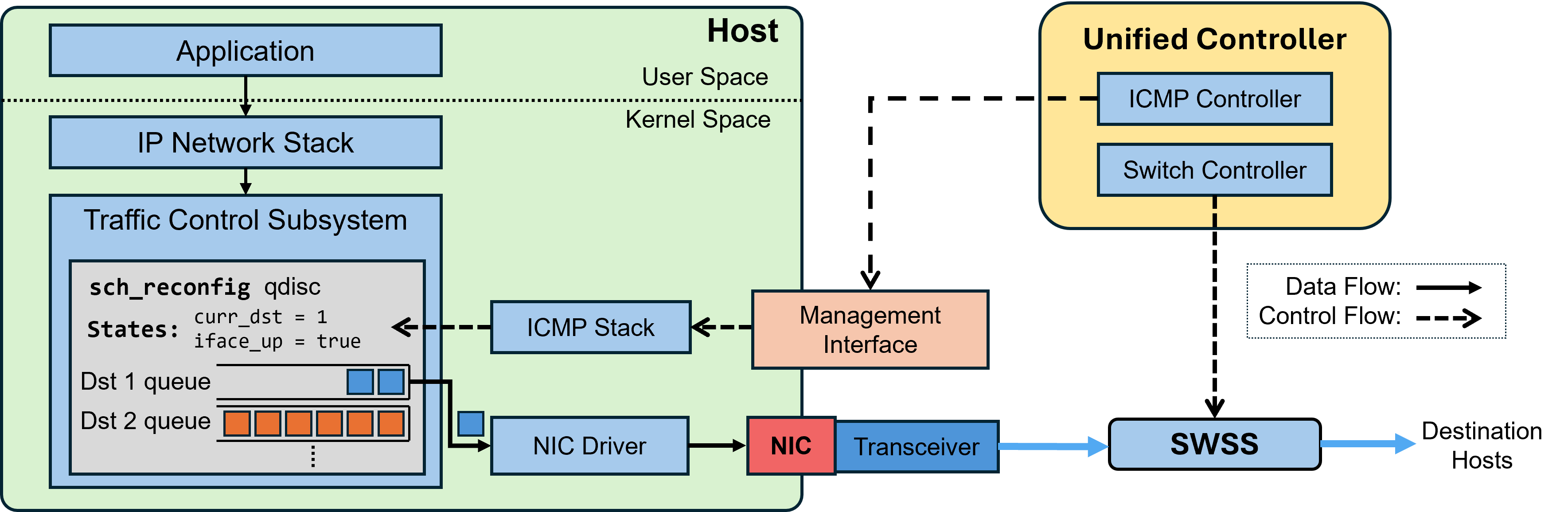}
    \caption{Software architecture of our host interface for direct-connect reconfigurable networks. On the host, application traffic descends through the IP network stack into the \texttt{sch\_reconfig} qdisc, which maintains the reconfiguration state (\texttt{curr\_dst}, \texttt{iface\_up}) and a per-destination queue, dispatching packets to the NIC driver only for the currently connected destination. The unified controller drives reconfiguration on both ends. Its switch controller thermally tunes the SWSS, while its ICMP controller signals the affected hosts through the management interface and the in-kernel ICMP stack, which updates the qdisc state accordingly.}
    \label{fig:software_arch}
\end{figure}

To support our custom SWSS, we implement the first Linux kernel packet scheduler for SWSS-based direct-connect reconfigurable networks. Figure~\ref{fig:software_arch} illustrates the software architecture. The host interface relies on two components. First, an ICMP-based notifier that accepts notifications from the switch or its out-of-band manager about an imminent reconfiguration. Second, a reconfiguration-aware queuing discipline (qdisc), built on the Linux traffic control framework, that briefly holds packets whose destinations are not connected or while the network is reconfiguring.

\begin{figure}[h!]
    \centering
    \includegraphics[width=\columnwidth]{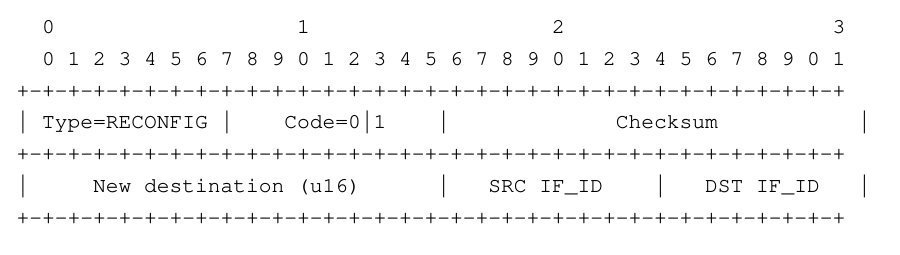}
    \caption{The ICMP message type for reconfiguration.}
    \label{fig:icmp}
\end{figure}

The ICMP subsystem notifies hosts explicitly of network state changes. We modify the Linux ICMP stack and define a new ICMP message type for this purpose, shown in Figure~\ref{fig:icmp}. Its header carries four fields. The ICMP message code marks the start and finish of reconfiguration (Code 0 and Code 1, respectively). The new destination is encoded as a 16-bit integer. The remaining two fields identify the affected interfaces on the source and destination hosts, which matters when a single host connects multiple interfaces to the SWSS. We pair this message with a unified controller that jointly orchestrates switch reconfiguration and host-side notification. Upon reconfiguration, the controller pushes pre-tuned voltage-control values to thermally tune the SWSS, acting as the switch controller. Simultaneously, it sends ICMP reconfiguration messages to the affected hosts, triggering them to register connection-destination changes and link-up/down events.

On the host, a new qdisc named \texttt{sch\_reconfig} tracks the current network state, manages traffic accordingly, and responds to reconfiguration events. Installed on each interface connected to the SWSS, the qdisc maintains two state variables: \texttt{curr\_dst} tracks the currently active destination and reflects the \texttt{New destination} field in the ICMP header; and \texttt{iface\_up} indicates whether the interface is up or is currently reconfiguring, as specified in the ICMP message code. Together, these variables cause the qdisc to block transmission whenever the destination is not the active connection or a reconfiguration blackout is in progress. The qdisc maintains a separate queue for each destination. A packet sent by the host is enqueued to its destination queue. The qdisc then continuously dispatches packets from the queue of the active destination while \textit{iface\_up} is set, and pauses all sending when the interface is down. Through the ICMP controller and its messages, \texttt{sch\_reconfig} becomes both reconfiguration- and destination-aware. Beyond the qdisc, we also modify the NIC driver to disable the administrative link-down state, since that state would flush the interface's routing and ARP tables, which must be preserved across reconfiguration.

\section{System Evaluation}
\label{sec:eval}

\begin{figure}[h!]
    \centering
    \includegraphics[width=0.9\linewidth]{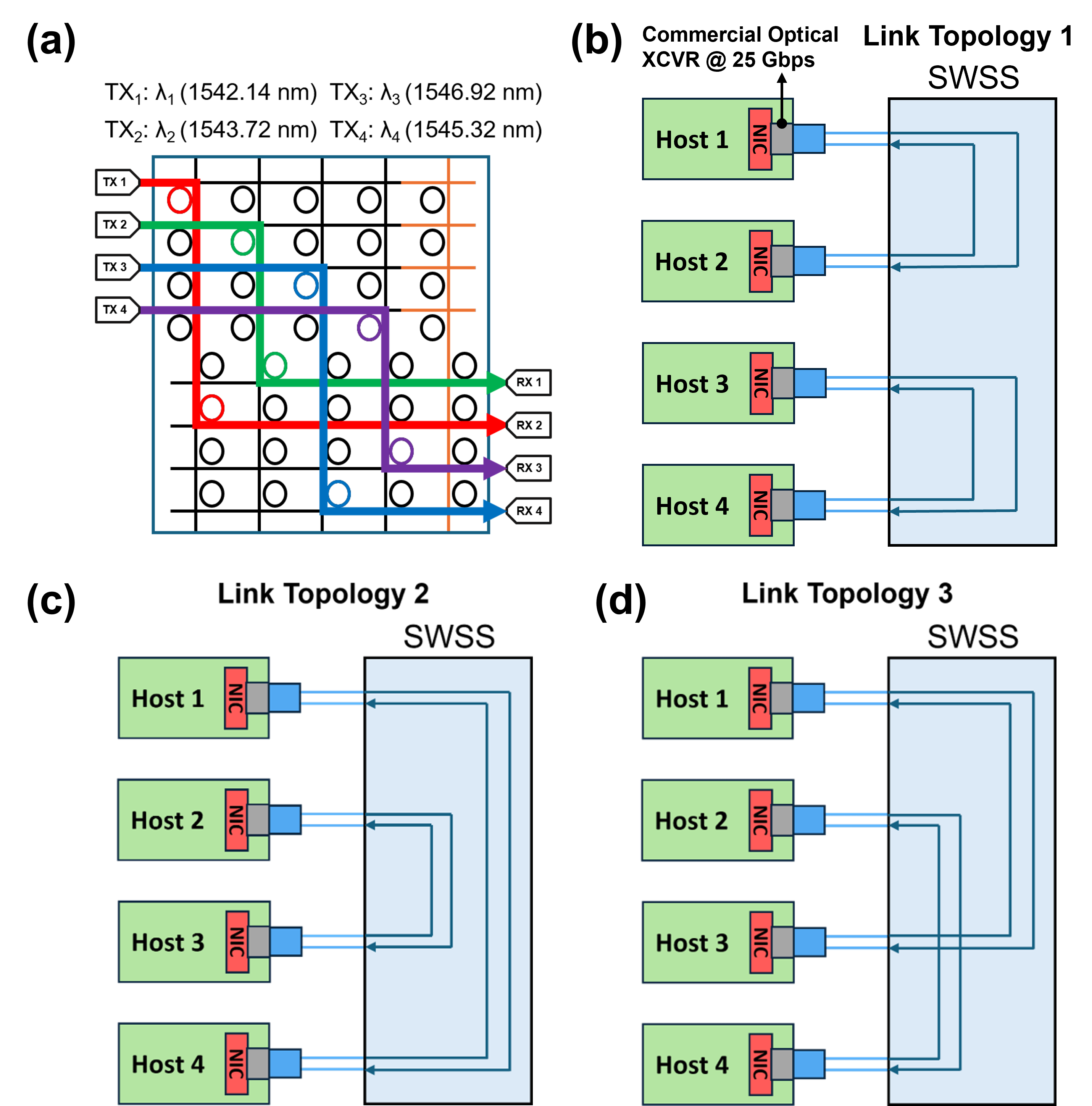}
    \caption{(a) Transmitter and receiver connections to SWSS along with assigned wavelength channels per host. (b-d) Link topologies 1 through 3 respectively.}
    \label{fig:setup}
\end{figure}

We demonstrate and evaluate our hardware-software integration for RADICAL with three end-to-end experiments. This 4-node implementation validates the feasibility of a directly connected reconfigurable optical fabric built with wavelength-selective switching, integrating a packaged $4\times4\times4\lambda$ SWSS, standard SFP28 transceivers, the Linux network stack, and real AI workloads. For our experimental setup, each host used a fixed-wavelength SFP28 transceiver on a discrete ITU channel: TX1/C44 at 1542.14 nm, TX2/C42 at 1543.73 nm, TX3/C38 at 1546.92 nm, and TX4/C40 at 1545.32 nm. These assignments are shown in Fig.~\ref{fig:setup}(a) with example routing and were chosen to demonstrate the spatial and wavelength selectivity of the SWSS. The device characterization measured spatial and wavelength crosstalk isolation across the calibrated SWSS configurations, with isolation exceeding 25 dB. In the system-level experiments, this isolation was sufficient to support stable data transfer and AI workload execution across repeated network reconfigurations. 

Prior work has quantified the potential performance benefits of optical and reconfigurable fabrics for distributed training through architecture-level studies. SiP-ML reports 1.3-9.1× faster training time than hierarchical electronic network fabrics for representative DNN models~\cite{Sipml}, while TopoOpt reports up to 3.4× faster DNN training than similar-cost Fat-Tree interconnects by co-optimizing network topology and parallelization strategy~\cite{Topoopt}. More recent work such as ACTINA compares OCS-based AI-training architectures against OCSFT-2L, which uses both EPSes and OCSes, and TPUv4-like OCS network architectures, reporting up to 1.75× better performance-power tradeoff compared with Fat-Tree and 3D-Torus-based OCS architectures~\cite{Actina}. Photonic-rail work such as Opus reports over 23× network power reduction and 4× cost savings while incurring less than 7\% training overhead at production-relevant OCS reconfiguration latencies~\cite{opus}. 

These results motivate reconfigurable optical fabrics as a promising direction for AI-training networks, while leaving open the complementary question of how wavelength-selective optical switching can be integrated into an end-to-end hardware-software system. RADICAL addresses this question by integrating a packaged SWSS with commodity servers, standard SFP28 transceivers, and the Linux network stack to support reconfiguration-aware operation. The multi-tenant experiments therefore demonstrate sustained system operation under SWSS-based topology changes using representative AI workloads. 

A direct application-level comparison against an EPS leaf-spine or static direct-connect baseline would require a separate optimized network implementation and training stack. Such an evaluation would shift the focus from integrating and validating an SWSS-based direct-connect fabric to optimizing and comparing full training-system implementations.

We evaluate RADICAL with three end-to-end experiments: (1) bulk data transfers over 10 minutes, (2) multitenant training of four instances of ResNet-18 across two link topologies, and (3) multitenant training of six instances of the nanoGPT language model across three link topologies. The first two experiments were demonstrated previously in the OFC '26 work~\cite{George_OFC}. The bulk-transfer and ResNet-18 experiments cycle between Link Topology 1 and Link Topology 2, while the nanoGPT experiment cycles through all three link topologies shown in Fig.~\ref{fig:setup}(b-d). For reference, Link Topology 1 connects GPU 1-GPU 2 and GPU 3-GPU 4; Link Topology 2 connects GPU 1-GPU 4 and GPU 2-GPU 3; and Link Topology 3 connects GPU 1-GPU 3 and GPU 2-GPU 4.

\subsection{Bulk Data Transfers}

\begin{figure}[h!]
    \centering
    \includegraphics[width=\linewidth]{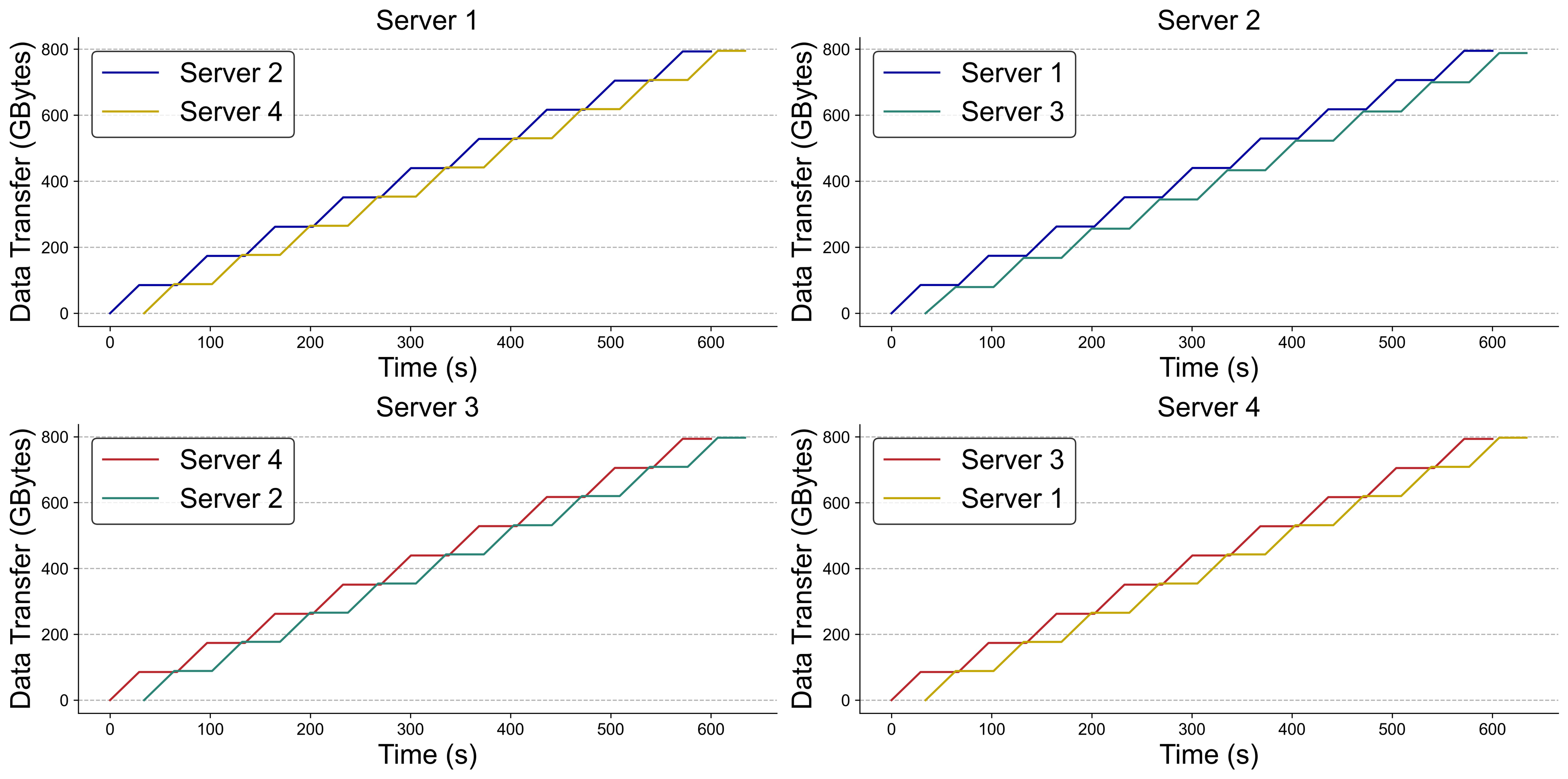}
    \caption{Cumulative data transferred (25 Gbps) of iPerf instances for four hosts where consecutive reconfigurations are illustrated every 30 seconds.}
    \label{fig:iperf}
\end{figure}

For the first end-to-end experiment, we chose to reconfigure between the first and second topology every 30 seconds. We generated traffic on the hosts by running simultaneous iPerf instances across each topology, four bidirectional pairs across the two configurations. As a result, we ran eight iPerf instances over the entire network as the SWSS reconfigured back and forth. Fig. \ref{fig:iperf} shows all eight instances of iPerf and the cumulative per-flow traffic. For example, server 1 has two flows each to different destination GPU servers for a particular link topology. 

After 10 minutes of running the iPerf instances, the network transmitted a total of 6.4 TBytes of data with only a single packet retransmission. Over 600 s, the measured transfer reached 6.4 TByte, or 85.3\% of the 7.50 TByte ideal for two active bidirectional 25 Gbps links. Traffic saturated the links between reconfigurations, and throughput recovered after each topology change. We also observe that the scheduler withholds packets when a destination is inactive and maintains a stable line-rate transmission across reconfigurations.

\subsection{Multitenant Training}

\begin{figure}[!h]
    \centering
    \includegraphics[width=0.7\linewidth]{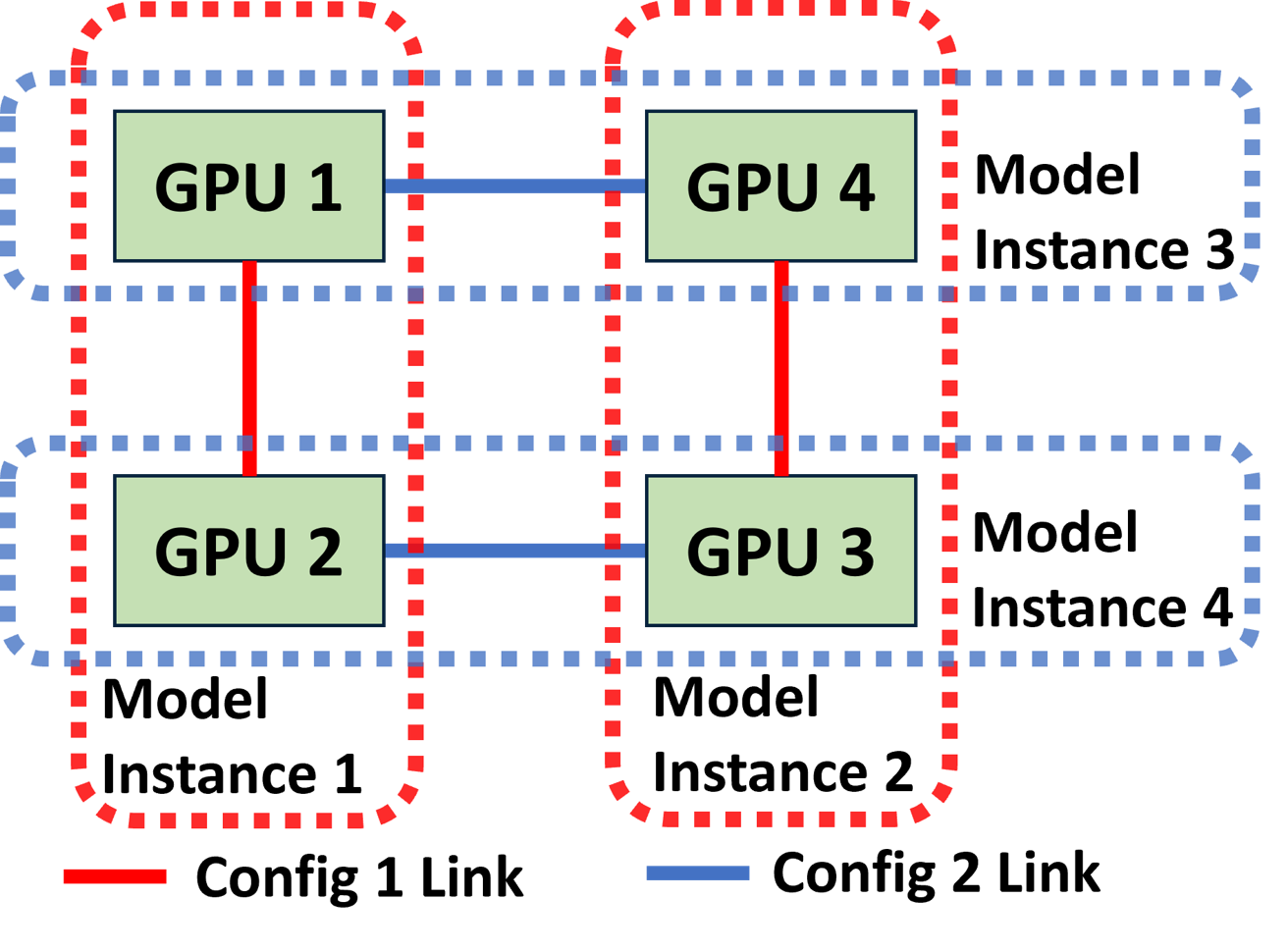}
    \caption{Multitenant training setup where each model was split across two GPUs that shared a link during a given link configuration.}
    \label{fig:model_split}
\end{figure}

The next two end-to-end experiments highlighted that RADICAL enables multitenant training across the cluster. In the first of these experiments \cite{George_OFC}, we trained four instances of ResNet-18 workloads across the network. We used the same reconfiguration schedule as the previous experiment, going between link topologies 1 and 2 every 30 seconds. Fig. \ref{fig:model_split} shows the four models were split across the network. Data parallelism was used to split a single model over two GPUs that share a link under a given configuration.

\begin{figure*}[h]
\centering

\begin{minipage}[c]{0.39\linewidth}
    \raggedright
    {\fontfamily{phv}\selectfont\textbf{(a)}}\\[0.2em]
    \centering
    \includegraphics[width=\linewidth]{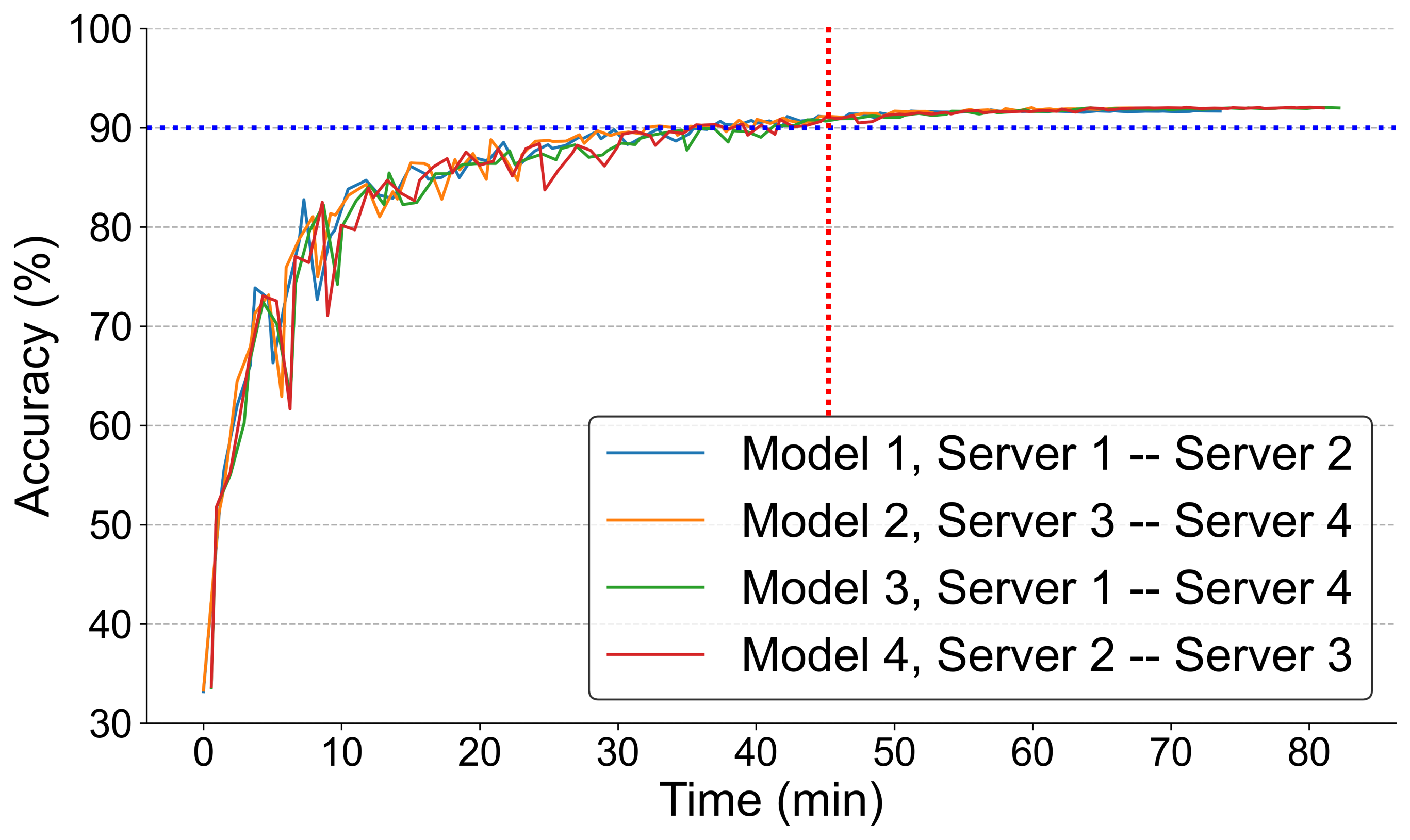}
\end{minipage}
\hfill
\begin{minipage}[c]{0.59\linewidth}
   \noindent
    {\fontfamily{phv}\selectfont\textbf{(b)}}%
    \hspace{0.42\linewidth}%
    {\fontfamily{phv}\selectfont\textbf{(c)}}\\[0.2em]
    \includegraphics[width=\linewidth]{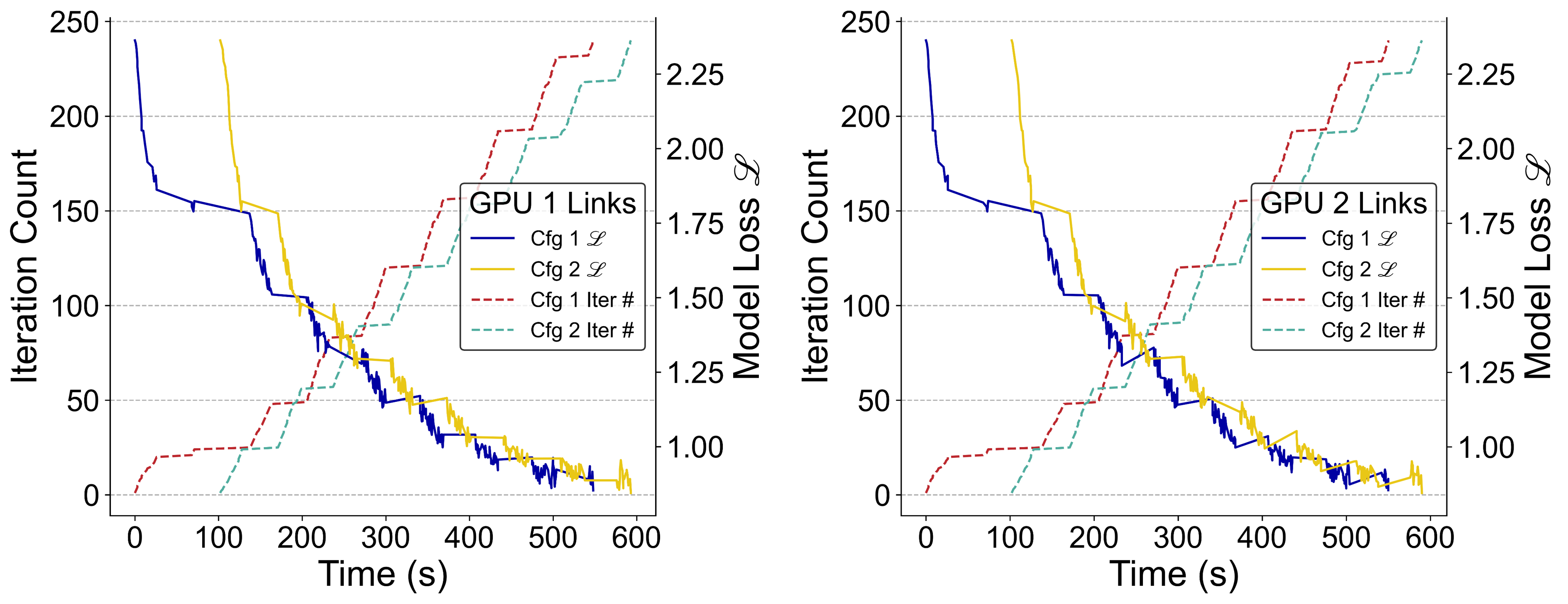}
\end{minipage}
\vspace{0.5em}

\caption{(a) Time-to-Accuracy plot for four ResNet-18 models converging concurrently. (b-c) Training loss and iteration count on GPU 1 and GPU 2 showing alternating convergence.}
\label{fig:timetoaccuracy}

\end{figure*}

The experiment proceeds as follows: two jobs are run on a single-link topology, then halted and buffered until the network reconfigures back to that topology. Fig. \ref{fig:timetoaccuracy}(a) shows all four instances reaching $90\%$ accuracy within 45 minutes of training. The entire training lasts 80 minutes in total, using CIFAR-10 for 100 epochs. The local batch size for these jobs was 1024. To illustrate that the network is reconfiguring every 30 seconds, we zoomed in on the iteration and model loss over the first 10 minutes of training in Figs. \ref{fig:timetoaccuracy}(b-c). These two plots show the per-iteration loss and iteration counts of the workloads across GPU-servers 1 and 2. As we would expect, the model loss decreases rapidly at the start of training as the iteration count increases. Both the time-to-accuracy and model loss metrics show that each ResNet-18 model is trained to convergence as the network cycles through reconfigurations. 

The final end-to-end experiment expands the complexity of the network by adding a third link topology, Fig. \ref{fig:setup}(d), while also tracking two additional models. The model used for this experiment is a compact 35.34M-parameter nanoGPT-style decoder-only Transformer trained on the enwik8 dataset using a 32k-token LLaMA 2 tokenizer. The model uses a 128-token context length, 6 layers, 8 attention heads, and 512-dimensional embeddings. Training was performed in FP32 with AdamW for 12,000 optimizer steps using a global batch size of 96 sequences. Similar to before, these models were parallelized across two GPU servers that shared a link across a given topology. Therefore, there were six instances of this nanoGPT model that were simultaneously trained across RADICAL.

\begin{figure*}[!h]
\centering
\includegraphics[width=\textwidth]{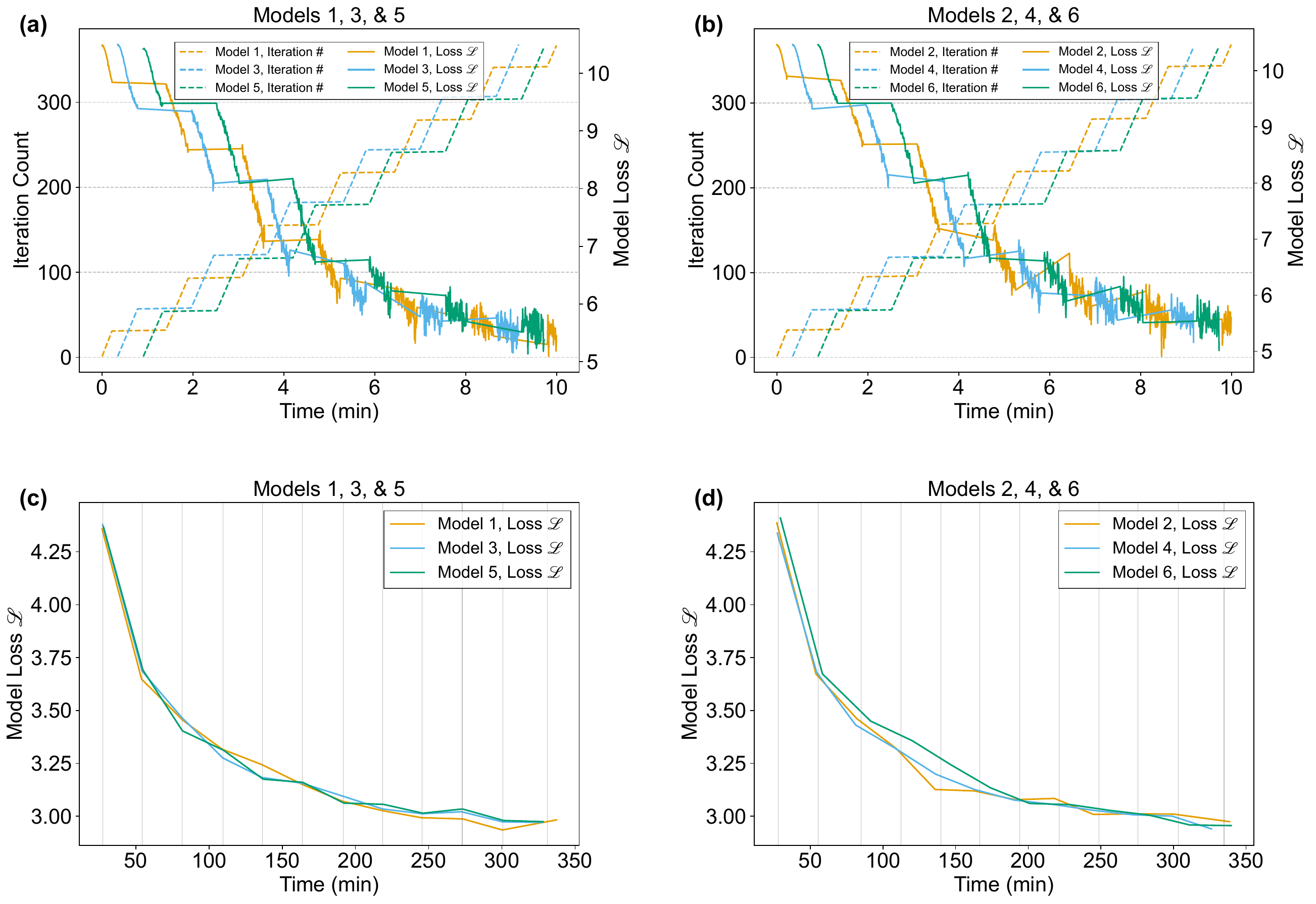}
\caption{(a) Per-iteration Model Loss and Iteration Count for Models 1, 3, and 5 for the first 10 minutes of training. (b) Per-iteration Model Loss and Iteration Count for Models 2, 4, and 6 for the first 10 minutes of training. (c) Per-epoch model loss versus time for Models 1, 3, and 5; vertical lines mark epoch boundaries. (d) Per-epoch model loss versus time for Models 2, 4, and 6; vertical lines mark epoch boundaries.}
\label{fig:threeconfigs}
\end{figure*}

We again use the same reconfiguration timing scheme for these experiments, every 30 seconds. The models trained during link topology 3 are designated as 5 and 6. Fig. \ref{fig:threeconfigs}(a-b) show the first 10 minutes of training for these models and illustrates the continuous reconfiguration of the system every 30 seconds. Here, we can also see the alternating and decreasing model loss across the three link configurations. We note that for each model, the per-iteration model loss decreases with iteration count. As expected, model loss decreases rapidly in the early stages of training for all three configurations. Fig. \ref{fig:threeconfigs}(c-d) show the per-epoch model loss for the more than 5 hours of training. For the nanoGPT workload, we report model loss rather than time-to-accuracy because it is intended to capture relative convergence behavior under each network configuration. The ResNet experiments use time-to-accuracy as a task-level metric since image-classification accuracy reaches a well-defined target during training. We also observe that in all cases the models continue to improve their loss with time and that the SWSS-interconnects stayed stable over this period. These demonstrations highlight our reconfigurable optical direct-connect architecture and full coordination between hardware and software planes.

These experiments demonstrate the functional capabilities required for an SWSS-based reconfigurable direct-connect AI cluster. The bulk data transfer experiments validate the reconfiguration-aware operation of the host network interface by showing alternating traffic flows that correspond to reconfiguration phases. Each link resumes line-rate transmission when the physical link is re-established for that period of time. The two multitenant ResNet-18 and nanoGPT training experiments highlight real AI workloads being trained correctly while the network continuously reconfigures and for an extended runtime. These results illustrate the feasibility of SWSS-based interconnects through a full hardware and software integration. Rather than a comparison to electronically packet switched networks, this work highlights the deployment of a custom SiPh SWSS with a reconfiguration-aware Linux packet scheduler into a commercial GPU server and transceiver environment.  

\section{Discussion and Scalability}

The present RADICAL system demonstrates the feasibility of SWSS-based direct-connect fabric using four commercial GPU servers and standard pluggable transceivers. Scaling this approach requires increasing the spatial radix of the SWSS and the number of WDM channels per port. Scaling can also be achieved by composing multiple SWSS devices across network dimensions. In doing this, the device must also maintain an acceptable optical power budget and practical control complexity. Each SWSS building block must also maintain crosstalk isolation. The present $4\times4\times4\lambda$ SWSS was characterized for path-dependent insertion loss and crosstalk across its calibrated spatial-wavelength configurations, as shown in Fig. \ref{fig:device_chars}. These measurements establish the optical margin of the demonstrated switch. In larger or denser SWSS devices, path-dependent loss can grow, while resonance alignment and thermal interactions can increase calibration complexity and crosstalk sensitivity. 

Prior work has shown that scaling the SWSS fabric is feasible \cite{8lambda_ofc} and the path loss remains manageable \cite{8lambda} to be used in system-level demonstrations. Larger radix switches would require substantially more control signals which would not be supported by peripheral wirebonding. Thus, future implementations for larger switches would likely require  flip-chip bonding, 3D-integration, or co-packaging techniques \cite{co-pack} for denser control signal density and channel monitoring. These approaches can also improve control-signal routing, thermal control, and calibration stability in larger SWSS devices.

At the system level, RADICAL provides a foundation for coarse-grained optical network flexibility. This work uses a fixed reconfiguration strategy that illustrates how this system could leverage the determinism exhibited by AI training workloads. These patterns of collective communication can be used through topology-aware mappings and scheduling \cite{Topoopt, Actina}. Scaling beyond this would require tighter coordination between the host servers and the switch controller. As discussed above, RADICAL targets coarse-grained, workload-aware reconfiguration rather than packet- or burst-level switching, so the graph-coloring and switch-control states can be computed offline and invoked by the out-of-band control plane at workload or training-phase boundaries. Also, by increasing the total number of compute nodes and network interface ports in the system, it enables us to evaluate different interconnect topologies, routing strategies, and reconfiguration policies. Each of these would be essential benchmarks for future system deployment. 

The two dimensions of SWSS flexibility play distinct roles. Spatial tunability determines which host pairs are directly connected and therefore changes the network topology available to the workload. Wavelength tunability determines how optical bandwidth is allocated across wavelength channels on those spatial connections, enabling finer-grained bandwidth steering when hosts can transmit or receive on multiple selectable wavelengths. In the present RADICAL implementation, the hosts use fixed-wavelength SFP28 transceivers, so the system-level experiments primarily evaluate reconfigurable spatial connectivity over a fixed host wavelength grid. However, the experiments still exercise the wavelength-selective function of the SWSS: four distinct host wavelengths are routed through the fabric across multiple topology configurations and to each of the possible receivers. Fully evaluating dynamic wavelength allocation as an independent performance knob would require tunable or WDM host transceivers so that the host interface can dynamically select or generate multiple wavelength channels. Recent work on co-designed SiPh I/O and reconfigurable Kerr-comb-driven DWDM links provides a path toward this form of high-bandwidth WDM host connectivity~\cite{yuyang, Cullen}. Such host-side wavelength flexibility is complementary to the SWSS integration demonstrated in RADICAL and remains a direction for future systems.

\section{Conclusion}

This paper presents RADICAL, the reconfiguration-aware direct-connect AI cluster using spatial-and-wavelength selective switching. The full hardware and software integration of our custom SiPh SWSS with a reconfiguration-aware Linux packet scheduler enables high-bandwidth, flexible interconnects between compute units. RADICAL uses a $4\times4\times4\lambda$ SWSS that uses MRR-based filters to provide dynamic selectivity across WDM channels with good crosstalk isolation between all wavelength and spatial ports and is capable of routing error-free WDM signals modulated at $25$ Gbps across the fabric. 

The system was evaluated by testing the dynamic control of both the SWSS device and the Linux packet scheduler to show dynamic link allocation between GPU-servers. Our system demonstrates bulk data transfers of 6.4 TBytes across 10 minutes, and multitenant training of ResNet-18 and nanoGPT models across continuous reconfigurations to convergence over an extended period of time. In both cases, the SWSS was implemented in a completely commercial data center environment, with the SWSS being the only specialized networking hardware which delivered the bandwidth and flexibility to train real AI workloads simultaneously. 

\section*{Funding}

This work is supported in part by the Laboratory for Physical Sciences (LPS) Research Initiative, and the Center for Ubiquitous Connectivity (CUbiC), and is sponsored by Semiconductor Research Corporation (SRC) and Defense Advanced Research Projects Agency (DARPA) under the JUMP 2.0 program.

\section*{Acknowledgments}

We thank HUBER+SUHNER Polatis for loaning the N-VST-192x192-LU1-DMHNS-800 OCS. We used this OCS for the initial development and validation of the Linux packet scheduler and ICMP controller, using the OCS's Telnet-based control interface as the switch controller.
 
Claude was used to assist with Python code for data visualization and post-processing. All interpretations and conclusions were performed and verified by the authors.

% Bibliography
\bibliography{swss_systems}

 \section*{Author Biographies}

\medskip

\noindent \textbf{Brett George} is a Ph.D. researcher in Prof. Keren Bergman's Lightwave Research Laboratory at Columbia University. He received his B.A. in Physics from the College of the Holy Cross, Worcester, MA, and his B.S. and M.S. in Electrical Engineering from Columbia University, New York, NY. His research interests include the design of silicon photonic devices such as WDM transceivers and switches and their system-level integration in data center networks.

\medskip

\noindent \textbf{Weiyang Wang} received the B.S. degrees in computer science and physics from the University of California San Diego, La Jolla, CA, USA, in 2020; the S.M. degree in computer science in 2022, and the Ph.D. degree in computer science in 2026 from the Massachusetts Institute of Technology, Cambridge, MA, USA.

He has held research internships at Google and Microsoft Research. In 2027, he will join New York University, New York, NY, USA, as an Assistant Professor of Computer Science. His research interests include adaptive network systems for machine learning workloads, with an emphasis on network architectures, host-adaptation of reconfigurable networks, and programmable datapaths.

\medskip

\noindent \textbf{Yuyang Wang} is an Assistant Professor at the University of Connecticut, Storrs, CT 06269, USA. His research interests include integrated photonics for communication and computing, electronic-photonic integration/packaging, and photonic design automation. Wang received his Ph.D. Degree in Electrical and Computer Engineering from the University of California, Santa Barbara. He was a Postdoctoral Research Scientist at Columbia University prior to joining UConn.

\medskip

\noindent \textbf{Zhenguo Wu} is a Research Scientist at NVIDIA Corporation at Santa Clara, California, 95051, USA. His research interests include optical interconnect and reconfigurable network design for machine learning systems. He received his B.S., M.S., and Ph.D. degrees in Electrical Engineering from Columbia University.

\medskip

\noindent \textbf{S.C. Mancas} is a Senior Researcher at the Laboratory for Physical
Sciences in Maryland. She received her Ph.D. in Applied Mathematics from the University of Central Florida, with dissertation research on dissipative solitons in the cubic–quintic complex Ginzburg–Landau equation, for which she received the UCF Outstanding Dissertation Award. With more than 20 years of experience as a mathematical physicist and computational scientist, she has an extensive record of peer-reviewed publications, conference proceedings, and technical reports to NASA and the National Science Foundation (NSF) and has presented her research at numerous national and international workshops and conferences. She also serves on the Editorial Advisory Board of Advances and Applications in Fluid Mechanics and as Co-Editor of Neural, Parallel, and Scientific Computations. Her research spans applied and mathematical physics, quantum information and computation, special functions, nonlinear dynamical systems, computational fluid dynamics, atmospheric modeling, and supersymmetric quantum mechanics. Her work emphasizes analytical and high-precision numerical methods, quantum algorithms, qubit and quantum-gate modeling, nonlinear evolution equations, Darboux transformations, and Schrödinger-type systems, with applications to quantum information, cryptography, optimization, fluid dynamics, and mathematical physics.

\medskip

\noindent \textbf{Xiang Meng} is an Associate Professor of Professional Practice in the Department of Electrical Engineering at Columbia University. His research focuses on photonic devices, coherent optical communication systems, and data-center technologies, with particular expertise in modeling, simulations, and cross-layer design spanning devices, systems, and networks. 

\medskip

\noindent \textbf{Manya Ghobadi} received her Ph.D. degree in computer science from the University of Toronto, Toronto, ON, Canada, in 2013.

Prior to joining the Massachusetts Institute of Technology, she spent several years at Microsoft Research and Google. She is currently an Associate Professor with the Department of Electrical Engineering and Computer Science, Massachusetts Institute of Technology, Cambridge, MA, USA. Her research interests include computer networks, with an emphasis on efficient systems for machine learning and high-performance cloud infrastructure.

Dr. Ghobadi has received the Sloan Research Fellowship in Computer Science, the ACM SIGCOMM Rising Star Award, the ACM-W Rising Star Award, the NSF CAREER Award, and the Optica Simmons Memorial Speakership Award. She has also received best paper awards at the Conference on Machine Learning and Systems (MLSys) and the IFIP Networking Conference, as well as the best dataset and best paper awards at the ACM Internet Measurement Conference (IMC).

\medskip

\noindent \textbf{Keren Bergman} is the Charles Batchelor Professor of Electrical Engineering at Columbia University and Faculty Director of the Columbia Nano Initiative. She earned her B.S. from Bucknell University and M.S. and Ph.D. in Electrical Engineering from MIT. At Columbia, she leads the Lightwave Research Laboratory at the intersection of computing and photonics. Bergman is a member of the American Academy of Arts and Sciences, a Fellow of IEEE and Optica, and a recipient of the IEEE Photonics Engineering Award and the Optica C.E.K. Mees Medal.

\medskip

\end{document}